\documentclass[twocolumn]{revtex4}
\usepackage{amsmath}
\usepackage{amssymb}
\usepackage{graphicx}
\usepackage{bm}
\usepackage{dcolumn}

\input{tcilatex}
\begin{document}

\title{Neutrons on a surface of liquid helium}
\author{P. D. Grigoriev}
\email{grigorev@itp.ac.ru}
\affiliation{L. D. Landau Institute for Theoretical Physics, Chernogolovka, Moscow
region, 142432, Russia;}
\author{O. Zimmer}
\email{zimmer@ill.fr}
\affiliation{Institut Laue-Langevin, BP 156, 6 rue Jules Horowitz, 38042 Grenoble Cedex
9, France}
\author{A. D. Grigoriev}
\affiliation{Samara State University, Samara, Russia}
\author{T. Ziman}
\affiliation{Institut Laue-Langevin, BP 156, 6 rue Jules Horowitz, 38042 Grenoble Cedex
9, France}

\begin{abstract}
We investigate the possibility of ultracold neutron (UCN) storage in quantum
states defined by the combined potentials of the Earth's gravity and the
neutron optical repulsion by a horizontal surface of liquid helium. We
analyse the stability of the lowest quantum state, which is most susceptible
to perturbations due to surface excitations, against scattering by helium
atoms in the vapor and by excitations of the liquid, comprised of ripplons,
phonons and surfons. This is an unusual scattering problem since the kinetic
energy of the neutron parallel to the surface may be much greater than the
binding energies perpendicular. The total scattering time constant of these
UCNs at $0.7$ K is found to exceed one hour, and rapidly increasing with
decreasing temperature. Such low scattering rates should enable
high-precision measurements of the scheme of discrete energy levels, thus
providing improved access to short-range gravity. The system might also be
useful for neutron $\beta $-decay experiments. We also sketch new
experimental concepts for level population and trapping of UCNs above a flat
horizontal mirror.
\end{abstract}

\date{\today }
\maketitle

\section{Introduction}

Slow neutrons play an important role in low-energy particle physics as a
tool and an object, in investigations of the free neutron's properties and
its interactions with known or hypothetic fields with high precision \cite%
{Dubbers/2011,Musolf/2008,Abele/2008}. A particular class of experiments
employs neutrons with energy lower than the neutron optical potential of
typical materials, i.e.\ in the order of up to $300$ neV. These so-called
ultracold neutrons (UCNs) can be emprisoned for many hundreds of seconds in
well-designed "neutron bottles". By virtue of the neutron magnetic moment of 
$60$ neV/T magnetic trapping is feasible too, and also the gravitational
interaction with a potential difference of $100$ neV per meter rise can play
its role in UCN storage and manipulation \cite{Golub/1991,Ignatovich/1990}.
Perhaps the most prominent application of UCNs is the search for a
non-vanishing electric dipole moment of the neutron. Finding a finite value
or just improving current limits is a way of investigating new mechanisms of
CP violation beyond the standard model's complex phase of the weak quark
mixing CKM matrix, and the matter-antimatter asymmetry in the universe \cite%
{Pospelov/2005}. Latest experimental results were published in Refs.\ \cite%
{Baker/2006,Serebrov/2014}. Similarly long standing are the experimental
efforts to determine the neutron lifetime with high accuracy. Its value
enters the calculations of weak reaction rates in big-bang nucleo-synthesis
and stellar fusion \cite{Coc/2007,Lopez/1999}, and it is also crucial for
deriving the weak axial-vector and vector coupling constants of the nucleon
needed to calculate many important semi-leptonic cross sections.

Discrete energy levels of UCNs in the Earth's gravitational field were
proposed by Lushikov and Frank in 1978 \cite{Luschikov/1978}, and
demonstrated experimentally in the past decade \cite%
{Nesvizhevsky/2002,Nesvizhevsky/2005,Westphal/2007}. Precise measurements of
the energy levels, that are not equi-distant, offer an interesting tool for
tests of various new scenarios of particle physics. The range of effects
investigated is determined by the characteristic size, several tens of
micrometers, with which the neutron wave functions are bound in one
dimension. Deviations from the Newton's gravity law at small distances can
for instance be interpreted as a signal of large extra dimensions at the
sub-millimeter scale \cite{Arkani-Hamed/1999,Antoniadis/2003} or as a hint
for dark-energy "chameleon" fields \cite{Brax/2011,Jenke/2014}. A recent
development, called gravity resonance spectroscopy (GRS), where transitions
between levels are induced by vibrating the mirror, has paved a way towards
sensitive tests of such scenarios \cite{Jenke/2011} (note, however, a strong
competition from atomic physics in the chameleon search \cite{Hamilton/2015}%
). A competing, alternative, method will employ oscillating magnetic field
gradients \cite{Kreuz/2009,Pignol/2014}. The GRS experiment described in
Ref.\ \cite{Jenke/2014} has already set stringent limits on chameleons. It
has also constrained axion-like particles, improving the result of an
analysis of the non-resonant gravity experiment described in Ref.\ \cite%
{Baessler/2007}. A method not relying on spatial quantum states of the
neutron employs spin precession of trapped UCNs close to a heavy mirror \cite%
{Zimmer/2010a}. The sensitivity of the GRS experiment was almost as good as
a first search of that latter type \cite{Serebrov/2010} and recently much
improved \cite{Afach/2015}, still with potential for large further gains in
sensitivity. For the gravity experiment too, a large gain is still to be
expected, notably once an adaptation of Ramsey's molecular beam technique of
separated oscillatory fields to GRS is implemented \cite{Abele/2010}. In
addition, a search for a non-zero neutron charge based on the latter
technique has been proposed \cite{Durstberger/2011}.

All current experiments on gravitational quantum states of the neutron
employ highly polished quartz mirrors. These are expensive, limited to sizes
of several tens of centimeters, and they have to be horizontally levelled by
some active means. In this respect, using a liquid surface as a mirror might
initiate a qualitatively new approach. On one hand it may furnish a remedy
to the aforementioned limitations. On the other hand, the interactions of
neutrons prepared in gravity states with excitations or structural
decorations of the liquid surface could enlarge the applications of these
states to investigations of the surface physics. The present article
provides a theoretical investigation of the possibility to store UCNs in the
lowest gravitational energy states on the liquid helium surface, by analysis
of scattering by helium atoms in the gas phase and by various excitations in
both the bulk and at the surface of the liquid helium. Obviously, a long
storage time constant is a necessary condition for conducting experiments
using a mirror made of this quantum liquid. A separate section sketches some
experimental concepts addressing issues arising in real studies employing
those neutrons, notably population, trapping and detection.

Properties of neutrons on the liquid helium surface are, in several aspects,
similar to those of electrons. The two-dimensional electron gas on a surface
of dielectric media has for many decades been a wide subject of research
(for reviews see, e.g., \cite{Shikin,Edelman,Monarkha}). In contrast to the
gravitational force in the neutron case, the electrons are attracted to the
boundary by the electric image forces through which they become localized in
the direction perpendicular to the surface. The surface of superfluid helium
has no solid defects (like impurities, dislocations, etc.) and offers a
unique opportunity to create an extremely pure 2D electron gas. The mobility
of electrons in this gas usually exceeds more than thousand times that of
electrons in 2D quantum wells in heterostructures. The system thus simulates
a solid-state 2D quantum well without disorder. Many fundamental properties
of a 2D electron gas have been studied with the help of electrons on the
surfaces of liquid helium. Various electronic quantum objects can be
experimentally realized on the liquid helium surface, such as quantum dots 
\cite{QuantumDot}, 1D electron wires \cite{QuantumWire}, quantum rings \cite%
{QuantumRing}, and others. The electrons on the liquid helium surface may
also serve for an experimental realization of a set of quantum bits with
very long decoherence time \cite{DykmanQC}. If neutrons can be made to rest
in surface states in sufficient densities we can hope for comparable studies
using neutrons rather than electrons, and possible new states of quantum
matter.

\section{Neutrons above a flat helium surface}

We consider a plane boundary between superfluid $^{4}$He (situated at
vertical coordinate $z<0$) and its saturated vapor ($z>0$). The interaction
of a neutron with a $^{4}$He atom with nuclear coordinate $\mathbf{R}_{i}$
can be expressed as a Fermi pseudo-potential given by 
\begin{equation}
V_{i}\left( \mathbf{r}\right) =U\delta ^{(3)}\left( \mathbf{r}-\mathbf{R}%
_{i}\right) \equiv \frac{2\pi \hbar ^{2}a_{\mathrm{He}}}{m}\delta
^{(3)}\left( \mathbf{r}-\mathbf{R}_{i}\right) ,  \label{Vi}
\end{equation}%
where $\delta ^{(3)}\left( \mathbf{r}\right) $ is the 3D Dirac delta
function. Substituting the bound coherent neutron scattering length $a_{%
\mathrm{He}}=3.26\times 10^{-13}$ cm of a $^{4}$He atom and the neutron mass 
$m=1.675\times 10^{-24}$ g, one obtains the value $U=1.36\times 10^{-42}$
erg cm$^{3}$ [$1$ erg = $10^{-7}$ J $\approx 6.25\times 10^{11}$ eV]. Within
the helium bulk the neutron interacts with a "forest" of $\delta $-function
potentials with volume concentration given by the particle density of $^{4}$%
He atoms, $n_{\mathrm{He}}\approx 21.8\ $nm$^{-3}$ at $T<2.18$ K. As a
result of the interference of a plane incident wave with the spherical
scattered waves from each $^{4}$He nucleus, neutron propagation in the bulk
can be described by a constant neutron optical potential given by the
spatially averaged pseudo-potentials of the helium atoms in a volume $\Omega 
$ containing many atoms, i.e. 
\begin{eqnarray}
V_{0} &=&\Omega ^{-1}\int_{\Omega }\sum_{i}~V_{i}\left( \mathbf{r}\right)
d^{3}\mathbf{r}=Un_{\mathrm{He}}  \label{V0} \\
&\approx &2.965\times 10^{-20}\ \mathrm{erg}=18.5\ \mathrm{neV}.  \notag
\end{eqnarray}%
Above the $^{4}$He surface, neglecting interactions with the helium vapor
discussed further below, the neutron is exposed to the gravity potential, 
\begin{equation}
V\left( z\right) =mg\,z,  \label{Vg}
\end{equation}%
where $g=981$ cm/s$^{2}$ is the acceleration of free fall (gravitational
acceleration at Earth's surface), giving the gravity force for the neutron, $%
mg=1.643\times 10^{-21}\ \mathrm{erg/cm}=1.026$ neV/cm.

\begin{figure}[tb]
\includegraphics[width=0.49\textwidth]{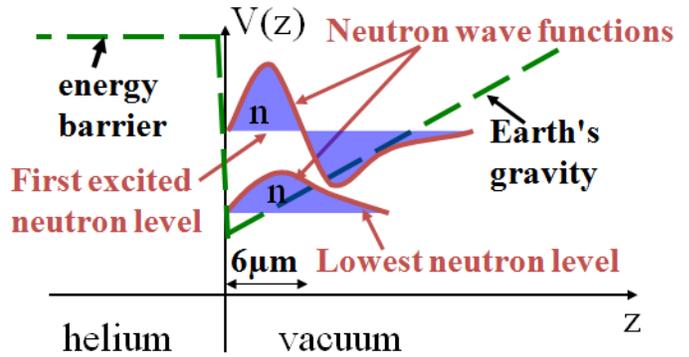}
\caption{Schematic representation of the vertical potential and the first
two states of a neutron above a horizontal mirror of liquid helium. }
\label{FigPot}
\end{figure}

One can easily solve the one-dimensional Schr\"{o}dinger equation for a
neutron in the potential given by Eqs.\ (\ref{V0}) and (\ref{Vg}) and shown
in Fig.\ \ref{FigPot}. The corresponding Hamiltonian is given by 
\begin{equation}
\hat{H}_{0}=-\frac{\hbar ^{2}\hat{\Delta}}{2m}+mgz+V_{0}\theta \left(
-z\right) ,  \label{H0}
\end{equation}%
where $\hat{\Delta}=\boldsymbol{\nabla }^{2}$ is the Laplace operator, $%
\boldsymbol{\nabla }$ is the gradient operator, and $\theta \left( x\right) $
is the step function [$\theta \left( x\right) =1$ for $x>0$ and $\theta
\left( x\right) =0$ for $x\leq 0$]. The $x$, $y$ and $z$ coordinates
separate in this equation, and the neutron wave function is given by a
product%
\begin{equation}
\psi \left( \mathbf{r}\right) =\psi _{\parallel }\left( \mathbf{r}%
_{||}\right) \,\psi _{\perp }\left( z\right) ,  \label{psiN}
\end{equation}%
where $\mathbf{r}_{||}=\left( x,y\right) $ is the 2D coordinate vector along
the surface, while $\mathbf{r}=\left( x,y,z\right) $ stands for the 3D
coordinate vector. In the $x$-$y$ plane, the neutron wave function is given
by a normalized plane wave, 
\begin{equation}
\psi _{\parallel }\left( \mathbf{r}_{||}\right) =S^{-1/2}\exp \left( i%
\mathbf{p}_{||}\mathbf{r}_{||}/\hbar \right) ,  \label{psiPar}
\end{equation}%
where $S$ is the He surface area, and $\mathbf{p}_{||}$ is the 2D neutron
momentum along the surface. For $z<0$ we can neglect the weak gravitational
potential given by Eq.\ (\ref{Vg}) as compared to the much stronger
potential wall given by Eq.\ (\ref{V0}). The $z$-dependent part of the
neutron wave function in this region is then approximately given by 
\begin{equation}
\psi _{\perp }\left( z\right) =\psi _{\perp }\left( 0\right) \exp \left(
\kappa z\right) \qquad \left( z<0,\ E_{\perp }<V_{0}\right) ,  \label{psi1}
\end{equation}%
where $\kappa =ik_{\perp }=\sqrt{2m\left( V_{0}-E_{\perp }\right) }/\hbar $,
and $E_{\perp }$ is the neutron kinetic energy along the $z$-axis. For $%
E_{\perp }\ll V_{0}$, $\kappa _{0}\equiv \sqrt{2mV_{0}}/\hbar \approx
2.4\times 10^{5}$ cm$^{-1}$, i.e.\ the neutron penetration depth into the
liquid helium is%
\begin{equation}
\kappa _{0}^{-1}\approx 33\ \mathrm{nm}.  \label{kappa_0}
\end{equation}%
For $z>0$, the neutron wave function is given by 
\begin{equation}
\psi _{\perp }\left( z\right) =C\,\mathrm{Ai}\left[ \left( z-E_{\perp
}/mg\right) /z_{0}\right] \qquad \left( z>0\right) ,  \label{Ai}
\end{equation}%
where $C$ is a normalization coefficient, $\mathrm{Ai}\left( x\right) $\ is
the Airy function, and 
\begin{equation}
z_{0}\equiv \left( \hbar ^{2}/2m^{2}g\right) ^{1/3}=5.87\ \mathrm{\mu m}
\label{z0}
\end{equation}%
is a characteristic length scale of the neutron wave function in low energy
states. For $z>E_{\perp }/mg$ the wave function given in Eq.\ (\ref{Ai})
decreases exponentially. For $E_{\perp }<V_{0}$ the energy spectrum of the $%
z $-axis motion of a neutron above the liquid-helium surface is quantized.
The eigenvalues of energy are given by the boundary condition at $z=0$, i.e. 
\begin{equation}
\frac{\psi _{\perp }^{\prime }\left( 0\right) }{\psi _{\perp }\left(
0\right) }=\kappa =\frac{\sqrt{2m\left( V_{0}-E_{\perp }\right) }}{\hbar }=%
\frac{\mathrm{Ai}^{\prime }\left( -u\right) }{z_{0}\mathrm{Ai}\left(
-u\right) },  \label{EqE}
\end{equation}%
where the prime indicates the derivative, and%
\begin{equation}
u\equiv E_{\perp }/mgz_{0}.
\end{equation}%
Equation (\ref{EqE}) for finite $V_{0}$ can be solved only numerically. A
characteristic scale of separations between lowest energy levels is given by 
$mgz_{0}=\left( \hbar ^{2}mg^{2}/2\right) ^{1/3}=0.96\times 10^{-24}\ 
\mathrm{erg}=0.6$ peV.

In the limit $V_{0}\rightarrow \infty $ the energy levels are given by 
\begin{equation}
E_{n}=mgz_{0}\alpha _{n+1},~n=0,1,...,  \label{En}
\end{equation}%
where $-\alpha _{n+1}$ are the zeros of the Airy function,%
\begin{equation}
\alpha _{1}=2.338,\quad \alpha _{2}=4.088,\quad \alpha _{3}=5.521,\quad
\alpha _{4}=6.787.  \label{alpha_i}
\end{equation}%
For $n\gg 1$, 
\begin{equation}
\alpha _{n}\approx \left( 3\pi n/2\right) ^{2/3}-\left( \pi ^{2}/96n\right)
^{1/3}.  \label{alphan}
\end{equation}%
Note that this expression provides a highly accurate approximation even for
small $n$: for $\alpha _{1}$ the error is only $\left\vert \Delta \alpha
/\alpha \right\vert \approx 0.0018$, and further decreases for $n>1$, e.g., $%
\left\vert \Delta \alpha /\alpha \right\vert \approx 0.0005$ for $\alpha
_{2} $.

For finite $V_{0}$, Eq.\ (\ref{EqE}) can be rewritten as 
\begin{equation}
\mathrm{Ai}^{\prime }\left( -u\right) =\mathrm{Ai}\left( -u\right) \sqrt{%
\eta _{0}-u},  \label{Equ}
\end{equation}%
using the dimensionless constant 
\begin{equation}
\eta _{0}\equiv V_{0}/mgz_{0}=3.078\times 10^{4},  \label{eta0}
\end{equation}%
evaluated on the r.h.s. for $V_{0}$ of superfluid He. Eqs.\ (\ref{Equ}) and (%
\ref{eta0}) give the following values of $\alpha _{n+1}^{\mathrm{He}%
}=E_{n}/mgz_{0}$ of the discrete energy of a neutron above the He mirror:%
\begin{equation}
\alpha _{1}^{\mathrm{He}}=2.332,\quad \alpha _{2}^{\mathrm{He}}=4.082,\quad
\alpha _{3}^{\mathrm{He}}=5.515,\quad \alpha _{4}^{\mathrm{He}}=6.781.
\end{equation}%
These values are indeed very close to the values $\alpha _{n}$ for $%
V_{0}\rightarrow \infty $ [compare Eq.\ (\ref{alpha_i})], because $\eta
_{0}\gg 1$. The neutron wave functions above liquid He are also very close
to those for $V_{0}\rightarrow \infty $, except for a region close to $z=0$,
where they acquires values $\psi _{\perp n}\left( 0\right) =C_{n}\,\mathrm{Ai%
}\left( -\alpha _{n+1}^{\mathrm{He}}\right) $ which are small but finite,
since $\alpha _{n+1}^{\mathrm{He}}\approx \alpha _{n+1}$. For the $n$-th
energy level the normalization coefficient is given by 
\begin{eqnarray}
\frac{1}{C_{n}^{2}} &=&\int_{0}^{\infty }dz\left[ \left\vert \mathrm{Ai}%
\left( \frac{z}{z_{0}}-\alpha _{n+1}^{\mathrm{He}}\right) \right\vert
^{2}+\left\vert \mathrm{Ai}\left( -\alpha _{n+1}^{\mathrm{He}}\right)
\right\vert ^{2}e^{-2\kappa z}\right]  \notag \\
&=&z_{0}\int_{0}^{\infty }dx\left\vert \mathrm{Ai}\left( x-\alpha _{n+1}^{%
\mathrm{He}}\right) \right\vert ^{2}+\frac{\left\vert \mathrm{Ai}\left(
-\alpha _{n+1}^{\mathrm{He}}\right) \right\vert ^{2}}{2\kappa }.  \label{Cn}
\end{eqnarray}%
The first three normalization coefficients are given by $C_{0}\approx 1.4261/%
\sqrt{z_{0}}\approx 59\ \mathrm{cm}^{-1/2}$, $C_{1}\approx 51.3\ \mathrm{cm}%
^{-1/2}$, $C_{2}\approx 47.8\ \mathrm{cm}^{-1/2}$. Using the values $\mathrm{%
Ai}\left( -\alpha _{1}^{\mathrm{He}}\right) \approx 0.004$, $\mathrm{Ai}%
\left( -\alpha _{2}^{\mathrm{He}}\right) \approx -0.0045$, and $\mathrm{Ai}%
\left( -\alpha _{3}^{\mathrm{He}}\right) \approx 0.0048$, this gives $\psi
_{\perp 0}\left( 0\right) \approx 0.236\ \mathrm{cm}^{-1/2}$, $\psi _{\perp
1}\left( 0\right) \approx -0.231\ \mathrm{cm}^{-1/2}$, $\psi _{\perp
2}\left( 0\right) \approx 0.23\ \mathrm{cm}^{-1/2}$, etc. These values will
be used further below.

In the subsequent sections we consider the stability of a neutron in a bound
surface state against various scattering processes. Note that we deal here
with a rather unfamiliar scattering problem in that the kinetic energy of
the neutron parallel to the surface may be many orders of magnitude greater
than the binding energies in the perpendicular direction. We calculate the
temperature-dependent scattering rates $w_{\mathrm{vap}}$, $w_{\mathrm{rip}}$
and $w_{\mathrm{sur}}$ due to $^{4}$He atoms in the vapour above the
surface, due to waves on the helium surface, called ripplons, and due to
excitations called surfons, respectively.

\section{ Scattering of neutrons in surface states by helium vapor}

$^{4}$He vapor atoms can be considered as point-like impurities with
interaction potential given by Eq.\ (\ref{Vi}). The momentum distribution of
the vapor atoms is given by the Bose distribution \cite{CommentBoltzmann}, 
\begin{equation}
N_{P}=\exp \left( \frac{\mu -E_{\mathrm{He}}}{k_{\mathrm{B}}T}\right) ,
\label{Np}
\end{equation}%
where $k_{\mathrm{B}}=1.38\times 10^{-16}\ \mathrm{erg/K}$ is the Boltzmann
constant, $\mu /k_{\mathrm{B}}=-7.17\ \mathrm{K}$ is the chemical potential
of liquid $^{4}$He (evaporation energy of a $^{4}$He atom) for $T\rightarrow
0$, and $E_{\mathrm{He}}=\mathbf{P}^{2}/2M$ is the kinetic energy of a $^{4}$%
He atom with momentum $\mathbf{P}$. The total atom concentration in the
vapor is given by the integrated momentum distribution, i.e.%
\begin{equation}
N_{\mathrm{vap}}=\left( \frac{Mk_{\mathrm{B}}T}{2\pi \hbar ^{2}}\right)
^{3/2}\exp \left( \frac{\mu }{k_{\mathrm{B}}T}\right) ,  \label{Ng}
\end{equation}%
where $M=6.7\times 10^{-24}\ \mathrm{g}$ is the atomic $^{4}$He mass.
Because of the energy-momentum conservation, the neutron scattering rate
depends on the initial neutron state and on $^{4}$He atom momentum. We
assume the initial neutron state to be given by the lowest energy level
along the $z$-axis and by the momentum $\mathbf{p}_{||}=\left\{
p_{x},p_{y}\right\} $ parallel to the helium surface, corresponding to the
total neutron energy%
\begin{equation}
K=p_{||}^{2}/2m+E_{0}.  \label{K_n}
\end{equation}%
In typical experiments with UCN, $K\approx p_{||}^{2}/2m\sim 10^{-7}\ 
\mathrm{eV}\gg E_{0}\sim 10^{-12}\ \mathrm{eV}$. The typical initial
momentum $\mathbf{P}$ of a He atom is larger than the neutron momentum by
still more than one order of magnitude, because its average kinetic energy $%
\bar{E}_{\mathrm{He}}=\left( 3/2\right) k_{\mathrm{B}}T\sim 10^{-4}$ eV.
Below [see Eq.\ (\ref{wp2})] we will see that, as $p_{||}\rightarrow 0$, the
neutron-He scattering rate $w_{\mathrm{vap}}$ remains finite, and we can
neglect $p_{||}/P\ll 1$ in the calculation of this rate.

\subsection{Matrix elements}

The initial neutron wave function is given by Eqs.\ (\ref{psiN}), (\ref%
{psiPar}) and (\ref{Ai}). Since the typical neutron out-of-plane kinetic
energy after scattering by a He atom of the vapor is much larger than $E_{0}$%
, and mostly even larger than $V_{0}$ given in Eq.\ (\ref{V0}), the final
neutron wave function is close to the three-dimensional plane wave with
momentum $\mathbf{p}^{\prime }$, normalized to one particle in the whole
volume $V$: 
\begin{equation}
\psi ^{\prime }=\exp \left( i\mathbf{p}^{\prime }\mathbf{r}/\hbar \right) /%
\sqrt{V}.  \label{psiNf}
\end{equation}%
The initial He-atom wave function is$\ \Psi =\exp \left( i\mathbf{Pr}/\hbar
\right) $, and the final He wave function is $\Psi ^{\prime }=\exp \left( i%
\mathbf{P}^{\prime }\mathbf{r}/\hbar \right) $. The matrix element of the
interaction potential (\ref{Vi}) is given by 
\begin{eqnarray}
T_{\mathrm{if}} &=&\int d^{3}\mathbf{r}\boldsymbol{\,}\psi _{\perp 0}\left(
z\right) \psi _{\parallel }\left( \mathbf{r}_{||}\right) \psi ^{\prime
}\left( \mathbf{r}\right) \times  \notag \\
&&\int d^{3}\mathbf{R}\exp \left( \frac{i\left( \mathbf{P}-\mathbf{P}%
^{\prime }\right) \boldsymbol{\mathbf{R}}}{\hbar }\right) U\,\delta
^{(3)}\left( \mathbf{r}-\mathbf{R}\right)  \notag \\
&=&U\,\int \frac{d^{3}\mathbf{R}}{\sqrt{SV}}\psi _{\perp 0}\left( z_{\mathrm{%
He}}\right) \exp \left( \frac{i\Delta \mathbf{P}_{\mathrm{tot}}\boldsymbol{%
\mathbf{R}}}{\hbar }\right) ,  \label{T1}
\end{eqnarray}%
where $\Delta \mathbf{P}_{\mathrm{tot}}\approx \mathbf{P}-\mathbf{P}^{\prime
}+\mathbf{p}_{||}-\mathbf{p}^{\prime }$ is the change of total momentum, $%
\mathbf{R}$ is the coordinate of the He nucleus, and $\psi _{\perp 0}\left(
z\right) =C_{0}\,\mathrm{Ai}\left[ \left( z-E_{0}/mg\right) /z_{0}\right] $
according to Eq.\ (\ref{Ai}). Introducing $\tilde{u}\equiv z/z_{0}$ and
performing the integration over $d^{2}\mathbf{R}_{\parallel }$ in Eq.\ (\ref%
{T1}) using the identity%
\begin{equation}
\int d^{2}\mathbf{R}_{\parallel }\exp \left( i\Delta \mathbf{P}_{\mathrm{tot}%
\parallel }\boldsymbol{\mathbf{R}}_{\parallel }/\hbar \right) =\left( 2\pi
\hbar \right) ^{2}\delta ^{(2)}\left( \Delta \mathbf{P}_{\mathrm{tot}%
||}\right) ,  \label{delta1}
\end{equation}%
one can rewrite $T_{\mathrm{if}}$ as%
\begin{equation}
T_{\mathrm{if}}=\frac{U\,\left( 2\pi \hbar \right) ^{2}\delta ^{(2)}\left(
\Delta \mathbf{P}_{\mathrm{tot}||}\right) }{\sqrt{SV/z_{0}}}I,  \label{Tif1}
\end{equation}%
where the remaining integral is%
\begin{equation*}
I\equiv 1.4261\int_{0}^{\infty }d\tilde{u}\,\,\mathrm{Ai}\left( \tilde{u}%
-\alpha _{1}\right) \exp \left( \frac{i\Delta p_{z}\boldsymbol{\,}\tilde{u}}{%
\hbar /z_{0}}\right) .
\end{equation*}%
We calculate this integral approximately by replacing the normalized Airy
function $f\left( \tilde{u}\right) \equiv 1.4261\mathrm{Ai}\left( \tilde{u}%
-\alpha _{1}\right) $ by a simpler form, also normalized, that is a close
approximation, i.e.$\ f\left( \tilde{u}\right) \approx \exp \left[ -\left( 
\tilde{u}-\tilde{u}_{0}\right) ^{2}/2\right] /\pi ^{1/4}$, where $\tilde{u}%
_{0}\approx 1.5$. Then,%
\begin{eqnarray}
I &=&\int_{0}^{\infty }d\tilde{u}\,\,f\left( \tilde{u}\right) \exp \left( 
\frac{i\Delta p_{z}\boldsymbol{\,}\tilde{u}}{\hbar /z_{0}}\right)  \notag \\
&\approx &\int_{-\infty }^{\infty }\frac{d\tilde{u}\,}{\pi ^{1/4}}\,\exp %
\left[ \frac{-\left( \tilde{u}-\tilde{u}_{0}\right) ^{2}}{2}+\frac{i\Delta
p_{z}\boldsymbol{\,}\tilde{u}}{\hbar /z_{0}}\right]  \notag \\
&=&\pi ^{1/4}\sqrt{2}\exp \left[ \frac{i\Delta p_{z}\boldsymbol{\,}\tilde{u}%
_{0}}{\hbar /z_{0}}-\frac{1}{2}\left( \frac{\Delta p_{z}\boldsymbol{\,}}{%
\hbar /z_{0}}\right) ^{2}\right] .  \label{I}
\end{eqnarray}%
Below we need only the square of the absolute value of the matrix element $%
T_{\mathrm{if}}$. The square of the $\delta $-function in $\left\vert T_{%
\mathrm{if}}\right\vert ^{2}$\ should be treated as%
\begin{equation}
\left[ \left( 2\pi \hbar \right) ^{2}\delta ^{(2)}\left( \Delta \mathbf{P}_{%
\mathrm{tot}||}\right) \right] ^{2}=S\left( 2\pi \hbar \right) ^{2}\delta
^{(2)}\left( \Delta \mathbf{P}_{\mathrm{tot}||}\right) ,  \label{delta2}
\end{equation}%
because it comes from the extra integration over the coordinate $\mathbf{r}%
_{i||}$: $\int d^{2}\mathbf{r}_{i\parallel }=S$. Indeed, substituting Eq.\ (%
\ref{delta1}) to the l.h.s. of Eq.\ (\ref{delta2}) we obtain \ 
\begin{gather*}
\left( 2\pi \hbar \right) ^{2}\delta ^{(2)}\left( \Delta \mathbf{P}_{\mathrm{%
tot}||}\right) \int d^{2}\mathbf{r}_{i\parallel }\exp \left( \frac{i\Delta 
\mathbf{P}_{\mathrm{tot}\parallel }\mathbf{r}_{i\parallel }}{\hbar }\right) =
\\
=\left( 2\pi \hbar \right) ^{2}\delta ^{(2)}\left( \Delta \mathbf{P}_{%
\mathrm{tot}||}\right) \int d^{2}\mathbf{r}_{i\parallel }=S\left( 2\pi \hbar
\right) ^{2}\delta ^{(2)}\left( \Delta \mathbf{P}_{\mathrm{tot}||}\right) .
\end{gather*}%
Substituting Eq.\ (\ref{I}) to Eq.\ (\ref{Tif1}) and using Eq.\ (\ref{delta2}%
), we obtain 
\begin{equation*}
\left\vert T_{\mathrm{if}}\right\vert ^{2}\approx \frac{U^{2}\,\left( 2\pi
\hbar \right) ^{3}\delta ^{(2)}\left( \Delta \mathbf{P}_{\mathrm{tot}%
||}\right) }{V\sqrt{\pi }\hbar /z_{0}}\exp \left[ -\left( \frac{\Delta p_{z}%
\boldsymbol{\,}}{\hbar /z_{0}}\right) ^{2}\right] .
\end{equation*}%
Since $\hbar /z_{0}\ll P$, using the identity%
\begin{equation*}
\delta \left( x\right) =\lim_{\epsilon \rightarrow 0}\left[ \frac{1}{%
\epsilon \sqrt{\pi }}\exp \left( -\frac{x^{2}}{\epsilon ^{2}}\right) \right]
,
\end{equation*}%
we rewrite $\left\vert T_{\mathrm{if}}\right\vert ^{2}$\ as%
\begin{equation}
\left\vert T_{\mathrm{if}}\right\vert ^{2}\approx U^{2}\,\left( 2\pi \hbar
\right) ^{3}\delta ^{(3)}\left( \Delta \mathbf{P}_{\mathrm{tot}}\right) /V.
\label{Tif2}
\end{equation}

\subsection{Scattering rate}

The scattering rate of a neutron with initial in-plane momentum $\mathbf{p}%
_{||}$ by a He atom with initial momentum $\mathbf{P}$ is given by the
square of the matrix element (\ref{Tif2}) integrated over the final momenta $%
\mathbf{p}^{\prime }$ and $\mathbf{P}^{\prime }$ of the neutron and He atom,
respectively (Fermi's golden rule \cite{LL3}): 
\begin{equation}
w_{\mathbf{P}}=\frac{2\pi }{\hbar }\int \frac{d^{3}\mathbf{P}^{\prime }}{%
(2\pi \hbar )^{3}}\int \frac{Vd^{3}\mathbf{p}^{\prime }}{(2\pi \hbar )^{3}}%
\left\vert T_{\mathrm{if}}\right\vert ^{2}\delta \left( \varepsilon
-\varepsilon ^{\prime }\right) .  \label{w1}
\end{equation}%
Here $\varepsilon \approx P^{2}/2M$ and $\varepsilon ^{\prime }=P^{\prime
2}/2M+\left( \mathbf{p}-\mathbf{p}^{\prime }\right) ^{2}/2m$ are the initial
and final total energies of He-atom and neutron. The scattering rate is
approximately independent of the initial neutron momentum $\mathbf{p}_{||}$,
i.e., $w_{\mathbf{P}}\equiv w\left( \mathbf{p}_{||},\mathbf{P}\right)
\approx w\left( \mathbf{P}\right) $, because $p_{||}\ll P$ can be neglected
in Eq.\ (\ref{w1}). We now substitute Eq.\ (\ref{Tif2}) to Eq.\ (\ref{w1}).
The integration over $\mathbf{p}^{\prime }$ cancels $\delta ^{3}\left(
\Delta \mathbf{P}_{\mathrm{tot}}\right) $ in Eq.\ (\ref{Tif2}), where\ $%
\Delta \mathbf{P}_{\mathrm{tot}}\approx \mathbf{P}-\mathbf{P}^{\prime }-%
\mathbf{p}^{\prime }$. After the integration over the angle $\phi $ between $%
\mathbf{P}$ and $\mathbf{P}^{\prime }$ we obtain 
\begin{eqnarray}
w_{\mathbf{P}} &=&\int \frac{P^{\prime 2}dP^{\prime }}{2\pi \hbar ^{4}}\frac{%
mU^{2}}{PP^{\prime }}\times \,  \label{wp1} \\
&&\theta \left[ 2PP^{\prime }-\left\vert P^{2}\left( 1-\frac{m}{M}\right)
+P^{\prime 2}\left( 1+\frac{m}{M}\right) \right\vert \right] .  \notag
\end{eqnarray}%
Using $M\approx 4m$, from Eq.\ (\ref{wp1}) we obtain 
\begin{equation}
w_{\mathbf{P}}\approx \int \frac{P^{\prime 2}dP^{\prime }}{2\pi \hbar ^{4}}%
\frac{mU^{2}}{PP^{\prime }}\theta \left[ 8PP^{\prime }-3P^{2}-5P^{\prime 2}%
\right] .  \label{wp}
\end{equation}%
The integrand is nonzero when the inequality 
\begin{equation}
5P^{\prime 2}-8PP^{\prime }+3P^{2}<0  \label{Inequality}
\end{equation}%
is satisfied. The quadratic expression has two real roots, $P^{\prime
}=P\left( 4\pm 1\right) /5$, so that Eq.\ (\ref{Inequality}) is satisfied
for $3/5<P^{\prime }/P<1$, thus defining the range of integration in Eq.\ (%
\ref{wp}), i.e.%
\begin{equation}
w_{\mathbf{P}}=\int_{3P/5}^{P}\frac{P^{\prime }dP^{\prime }}{2\pi \hbar ^{4}}%
\frac{mU^{2}}{P}=\frac{U^{2}Pm}{2\pi \hbar ^{4}}\frac{8}{25}.  \label{wp2}
\end{equation}%
Finally, to obtain the total scattering rate as function of temperature one
has to integrate Eq.\ (\ref{wp2}) over the initial He-atom momentum $\mathbf{%
P}$, weighted with the distribution function $N_{P}$ of He vapor given by
Eq.\ (\ref{Np}), 
\begin{eqnarray*}
w_{\mathrm{vap}}\left( T\right) &=&\int \frac{d^{3}\mathbf{P}~N_{P}}{(2\pi
\hbar )^{3}}w_{\mathbf{P}} \\
&=&\int \frac{P^{2}dP^{2}~}{(2\pi \hbar )^{3}}\frac{U^{2}}{\hbar ^{4}}\frac{%
8m}{25}\exp \left( \frac{\mu -P^{2}/2M}{k_{\mathrm{B}}T}\right) .
\end{eqnarray*}%
Introducing the new dimensionless variable $P^{2}/2Mk_{\mathrm{B}}T$ and
performing the integration we obtain%
\begin{equation}
w_{\mathrm{vap}}\left( T\right) =\frac{\left( 2Mk_{\mathrm{B}}T\right) ^{2}}{%
(2\pi \hbar )^{3}}\frac{U^{2}}{\hbar ^{4}}\frac{8m}{25}\exp \left( \frac{\mu 
}{k_{\mathrm{B}}T}\right) .  \label{wv}
\end{equation}%
After substitution of Eq.\ (\ref{Vi}) one obtains%
\begin{equation}
w_{\mathrm{vap}}\left( T\right) =9.44\ \mathrm{s}^{-1}\times \left( T\left[ 
\mathrm{K}\right] \right) ^{2}\times \exp \left( \frac{-7.17}{T\left[ 
\mathrm{K}\right] }\right) .  \label{wv2}
\end{equation}%
Hence, $w_{\mathrm{vap}}\left( 1\ \mathrm{K}\right) \approx 0.007\ \mathrm{s}%
^{-1}=(138\ \mathrm{s})^{-1}$, and the estimated mean scattering time of a
neutron in the lowest level, as determined by He vapor only, is about $138$
s at $T=1$ K. Lowering the temperature diminishes the scattering rate more
rapid than exponentially, e.g., $w_{\mathrm{vap}}\left( T=0.8\ \mathrm{K}%
\right) \approx 7.74\times 10^{-4}\ \mathrm{s}^{-1}=\left( 21.53\text{ min}%
\right) ^{-1}$, and $w_{\mathrm{vap}}\left( T=0.7\ \mathrm{K}\right) \approx
1.6\times 10^{-4}$ s$^{-1}$. The break-even with neutron decay is thus
reached slightly above $0.8$ K.

\section{Scattering from surface waves}

\subsection{General information about ripplons}

A quantum of a surface wave (ripplon) with momentum $\mathbf{q}$ induces a
surface deformation along the $z$-axis, given by 
\begin{equation}
\xi (\mathbf{r}_{\parallel },t)=\xi _{0q}\sin \left( \mathbf{qr}_{\parallel
}-\omega _{q}t\right) .  \label{xi2}
\end{equation}%
The dispersion relation of surface waves is given by \cite{LL6,Shikin} 
\begin{equation}
\omega _{q}^{2}=\frac{\alpha }{\rho }\left( q^{2}+\varkappa ^{2}\right)
q\tanh \left( qd\right) ,  \label{2}
\end{equation}%
where $\alpha \approx 0.354$ dyn/cm is the surface tension coefficient of
superfluid $^{4}$He, $\rho \approx 0.145$ g/cm$^{3}$ is its mass density, $d$
is the depth of the helium bath above a horizontal bottom wall, and $%
\varkappa ^{2}=\left( g+f\right) \rho /\alpha $ with an additional force $%
f\propto d^{-4}$ due to the van-der-Waals attraction of helium to the bottom
wall. The ripplon amplitude $\xi _{0q}$ in Eq.\ (\ref{xi2}), normalized to
one ripplon per surface area $S$, is given by \cite{Shikin,SM,CommentSW} 
\begin{equation}
\xi _{0q}=\left( \frac{\hbar q\tanh \left( qd\right) }{2S\,\rho \omega _{q}}%
\right) ^{1/2}.  \label{xi0q}
\end{equation}%
For a helium bath (in fact already for a thick helium film), $\varkappa =%
\sqrt{g\rho /\alpha }\approx 20$ cm$^{-1}$. The thermal ripplons with energy 
$\hbar \omega _{q}\approx k_{\mathrm{B}}T\approx 0.5$ K have the wave number 
$q\approx 1.2$ nm$^{-1}$, for which holds $q\gg \varkappa $ and $qd\gg 1$.
Then the dispersion relation of ripplons is just the dispersion of capillary
waves: 
\begin{equation}
\omega _{q}\approx \sqrt{\alpha /\rho }q^{3/2},  \label{3}
\end{equation}%
and 
\begin{equation}
\xi _{0q}\approx \left( \frac{\hbar }{2S\,\sqrt{\rho \alpha q}}\right)
^{1/2}.  \label{xi0qa}
\end{equation}

\subsection{Interaction Hamiltonian}

To determine the influence of a periodic surface deformation on the neutron
quantum state on the surface we have to separate two limits. The first, 
\textit{adiabatic} limit appears when the surface oscillates so slowly that
the neutron wave function adjusts to the instantaneous surface profile. The
interaction potential in this limit is found in Appendix A, see Eq.\ (\ref%
{Hint0}), and can be rewritten as%
\begin{equation}
\hat{H}_{\mathrm{int}}=\xi (\mathbf{r}_{\parallel },t)\left\{ \left[ \frac{%
\left( \hat{p}_{\parallel }+\hat{p}_{q}\right) ^{2}-\hat{p}_{\parallel }^{2}%
}{2m}-\hbar \omega _{q}\right] \frac{\partial }{\partial z}+mg\right\} ,
\label{Hint}
\end{equation}%
where $\hat{p}_{\parallel }=-i\hbar \boldsymbol{\nabla }_{\parallel }$ and $%
\hat{p}_{q}=\hbar q$ are the momentum operators of the neutron and ripplon
along the surface, respectively. This interaction term generalizes Eq.\ (7)
of Ref. \cite{Abele/2010}, because it does not exclude coordinate-dependent
surface perturbations. The expression in the square brackets in Eq.\ (\ref%
{Hint}) is just the transfer of the total (neutron+ripplon) kinetic energy
to the final neutron kinetic energy along the $z$-axis.

The opposite, anti-adiabatic or \textit{diabatic} limit appears when the
surface oscillates much faster than the characteristic frequency of the
out-of-plane neutron motion, so that the neutron wave function does not
adjust to the instantaneous surface profile. In this limit a surface wave
affects the neutrons by creating an additional time- and
coordinate-dependent periodic potential 
\begin{equation}
V_{r}(\mathbf{r}_{\parallel },z)=\left\{ 
\begin{array}{c}
V_{0}~\text{at }0<z<\xi (\mathbf{r}_{\parallel },t)\text{ for }\xi (\mathbf{r%
}_{\parallel },t)>0 \\ 
-V_{0}\text{ at }\xi (\mathbf{r}_{\parallel },t)<z<0\text{ for }\xi (\mathbf{%
r}_{\parallel },t)<0%
\end{array}%
\right. .  \label{Vr}
\end{equation}%
The ripplon amplitude, given by Eqs.\ (\ref{xi0q}) or (\ref{xi0qa}), for any
reasonable value of $S$ is much less than the atomic scale and, even more,
than the typical scale of the neutron wave function, given by $\kappa
_{0}^{-1}\approx 33$ nm at $z<0$ [Eq.\ (\ref{kappa_0})]. Therefore, the
potential in Eq.\ (\ref{Vr}) can be approximated by 
\begin{equation}
V_{r}(\mathbf{r})\equiv V_{r}(\mathbf{r}_{\parallel },z)\approx V_{0}\xi (%
\mathbf{r}_{\parallel },t)\delta \left( z\right) .  \label{Vr1}
\end{equation}

\subsection{Crossover between adiabatic and diabatic limits and matrix
elements}

The diabatic-adiabatic\ crossover, corresponding to a change of the
ripplon-neutron interaction Hamiltonian from Eq.\ (\ref{Vr}) to Eq.\ (\ref%
{Hint}), must take place when the ripplon frequency $\omega _{q}$ and the
wave-vector $\mathbf{q}$ decrease. However, the estimate of the crossover
frequency $\omega _{q\mathrm{c}}$ and the description of the system in the
crossover regime is not a trivial problem. Similar problem appears in other
condensed-matter systems and requires a special theoretical study (see,
e.g., Refs. \cite{Eidel2013,Giamarchi2005,Dora2011,Feig1998,Ogawa1999}).

One may, naively, define the crossover as the region where the ripplon
frequency becomes comparable to the quasi-classical bouncing frequency of a
neutron in the ground level in $z$-direction, i.e. when the ratio $\hbar
\omega _{q}/E_{0}\sim 1$, where $E_{0}$ given by Eq.\ (\ref{En}). This
corresponds to the ripplon frequency 
\begin{equation}
\omega _{q\mathrm{c}}\sim E_{0}/\hbar =915\ \mathrm{s}^{-1},  \label{omc2}
\end{equation}%
and to the ripplon wave number 
\begin{equation}
q_{\mathrm{c}}\approx \left( \omega _{q\mathrm{c}}^{2}\rho /\alpha \right)
^{1/3}=70\ \mathrm{cm}^{-1}>\varkappa .  \label{qc2}
\end{equation}%
However, such an estimate of the diabatic-adiabatic\ crossover has an
important drawback: it does not depend on the value $V_{0}$ of the neutron
potential inside helium. Generally, we expect that for $V_{0}\rightarrow 0$
and for non-zero $q$ and $\omega _{q}$ one can always apply Eq.\ (\ref{Vr}),
and for $V_{0}\rightarrow \infty $ one can always apply Eq.\ (\ref{Hint}),
which contradicts Eq.\ (\ref{omc2}). The classical definition of the
diabatic-adiabatic\ crossover, given by Eqs.\ (\ref{omcCl}) and (\ref{qcCl})
in Appendix B,\ has the same drawback.

A rigorous analysis of the adiabatic/diabatic crossover should be based on
the solution of the Schr\"{o}dinger equation for a neutron in the
time-dependent potential given by Eqs.\ (\ref{Schrod}) and (\ref{H}). One
may approximately determine the criterion of adiabatic/diabatic crossover
from the variational principle to minimize the neutron energy. This approach
would be definitely correct for a time-independent potential. The
lowest-level out-of-plane neutron wave function is chosen to minimize the
neutron energy. In the adiabatic limit the energy loss is the kinetic and
gravitational energy from Eq.\ (\ref{Hint}), while in the diabatic limit it
is the potential energy from Eq.\ (\ref{Vr}). The first-order energy
correction is given by the diagonal matrix elements of these two interaction
potentials, or more precisely by the Hamiltonian in Eq.\ (\ref{H}). If these
diagonal matrix elements are nonzero, their comparison gives the crossover
frequency. If these matrix elements vanish in the first order in $\xi $, one
needs to calculate and compare the second-order corrections. Since $%
\int_{-\infty }^{\infty }dz\psi _{\perp 0}^{\ast }\left( z\right) \partial
\psi _{\perp 0}\left( z\right) /\partial z=0$, and $\int \xi (\mathbf{r}%
_{\parallel },t)d^{2}\mathbf{r}_{||}=0$, the first-order (in $\xi $)
diagonal matrix element of the adiabatic Hamiltonian in Eq.\ (\ref{Hint})
vanishes. So does the diagonal matrix element of the diabatic Hamiltonian in
Eq.\ (\ref{Vr}) in the first-order in $\xi $, if $\mathbf{p}_{||}\neq \hbar 
\mathbf{q}$. Hence, to calculate the crossover frequency one needs to
calculate the second-order energy corrections, which do not vanish. These
corrections are determined, in particular (but not only), by the matrix
elements of the neutron-ripplon interaction potentials in Eqs.\ (\ref{Hint})
and (\ref{Vr}). Therefore, for an estimate of the position (ripplon
frequency) of the adiabatic/diabatic crossover the comparison of the matrix
elements, given below, is more accurate than just the comparison of ripplon
frequency with $E_{0}$. As we will see, the final result of the
neutron-ripplon scattering rate is not sensitive to this crossover
frequency, because the main contribution to this scattering rate comes from
the ripplons with energy $\hbar \omega _{q}\sim V_{0}\gg \hbar \omega _{q%
\mathrm{c}}$, which corresponds to the far diabatic limit.

Therefore, a rough estimate of the crossover between diabatic and adiabatic
limits is given by the ripplon frequency when two interaction Hamiltonians,
given by Eqs.\ (\ref{Vr}) and (\ref{Hint}), become of the same order of
magnitude. More precisely, we compare their matrix elements for the neutron
transitions between lowest energy levels of their motion in $z$-direction.
Thus defined, the diabatic-adiabatic crossover depends on $V_{0}$ and meets
other general requirements, such as the adiabatic limit for $\omega
_{q},q\rightarrow 0$. The matrix element $T_{\mathrm{if}}$ of the diabatic
interaction potential in Eq.\ (\ref{Vr1}) for the transitions between two
neutron states with initial wave function $\psi _{\perp }\left( z\right) $
and final wave function $\psi _{\perp }^{\prime }\left( z\right) ,$ written
explicitly in Eqs.\ (\ref{psiN})-(\ref{Ai}), is given by 
\begin{equation*}
T_{\mathrm{if}}=\int \frac{d^{3}\mathbf{r}\boldsymbol{\,}}{S}\psi _{\perp
}\left( z\right) \psi _{\perp }^{\prime \ast }\left( z\right) \exp \left( i%
\mathbf{r}_{||}\frac{\mathbf{p}_{||}-\mathbf{p}_{\parallel }^{\prime }\,}{%
\hbar }\right) V_{0}\xi (\mathbf{r}_{\parallel })\delta \left( z\right) .
\end{equation*}%
The integral over $z$ cancels $\delta \left( z\right) $, while after
substituting Eq.\ (\ref{xi2}) the integration over $\mathbf{r}_{||}$ gives $%
\left( 2\pi \hbar \right) ^{2}\delta ^{(2)}\left( \Delta \mathbf{p}_{\mathrm{%
tot}||}\right) $, where $\Delta \mathbf{p}_{\mathrm{tot}||}=\hbar \mathbf{q}+%
\mathbf{p}_{||}-\mathbf{p}_{\parallel }^{\prime }$ is the change of the
total in-plane momentum of the ripplon+neutron system. As a result we obtain 
\begin{equation}
T_{\mathrm{if}}=V_{||}V_{0,n}\,.  \label{Tif}
\end{equation}%
The factor 
\begin{equation}
V_{||}=S^{-1}\left( 2\pi \hbar \right) ^{2}\delta ^{(2)}\left( \Delta 
\mathbf{p}_{\mathrm{tot}||}\right)  \label{Vp}
\end{equation}%
is due to the in-plane part $\psi _{\parallel }\left( \mathbf{r}_{||}\right) 
$ of the neutron wave function, given by Eq.\ (\ref{psiPar}), and 
\begin{equation}
V_{0,n}=V_{0}\xi _{0q}\psi _{\perp 0}^{\ast }\left( 0\right) \psi _{\perp
n}\left( 0\right)  \label{V0n}
\end{equation}%
comes from its out-of-plane part $\psi _{\perp }\left( z\right) $. The
squared modulus of the matrix element in Eq.\ (\ref{Tif}) follows as 
\begin{equation}
\left\vert T_{\mathrm{if}}\right\vert ^{2}=\frac{\left( 2\pi \hbar \right)
^{2}\delta ^{(2)}\left( \Delta \mathbf{p}_{\mathrm{tot}||}\right) }{S}%
\left\vert \psi _{\perp }\left( 0\right) \psi _{\perp }^{\prime \ast }\left(
0\right) V_{0}\xi _{0q}\right\vert ^{2},  \label{Tif2r}
\end{equation}%
where we again have used Eq.\ (\ref{delta2}).

The matrix elements of the adiabatic interaction potential in Eq.\ (\ref%
{Hint}) for $\xi (\mathbf{r}_{\parallel },t)$ given by Eq.\ (\ref{xi2}) are 
\begin{equation}
\hat{H}_{k,n}=V_{||}\xi _{0q}\left( E_{k}-E_{n}\right) Q_{k,n},  \label{Hkn}
\end{equation}%
where $V_{||}$ is again given by Eq.\ (\ref{Vp}) and%
\begin{equation}
Q_{k,n}=\int_{0}^{\infty }dz\psi _{\perp k}^{\ast }\left( z\right) \frac{%
d\psi _{\perp n}^{\ast }\left( z\right) }{dz}.  \label{Qkn}
\end{equation}%
The first values of $Q_{k,n}$ are given in Table I of Ref. \cite{Abele/2010}%
, e.g., $Q_{0,1}=0.09742\ \mathrm{\mu m}^{-1}$, $Q_{0,2}=-0.05355\ \mathrm{%
\mu m}^{-1}$, $Q_{0,3}=0.03831\ \mathrm{\mu m}^{-1}$, $Q_{0,4}=-0.0304\ 
\mathrm{\mu m}^{-1}$. Substituting these values we obtain for the first
levels the ratio $V_{0,n}/\hat{H}_{0,n}\approx 1$ at 
\begin{equation}
\omega _{q\mathrm{c}}\approx 10^{3}\ \mathrm{s}^{-1}.  \label{omc}
\end{equation}%
By chance, the diabatic-adiabatic crossover condition, defined as $%
V_{0,n}\sim \hat{H}_{0,n}$, is close to the value quoted in Eq.\ (\ref{omc2}%
).

Below, we consider mainly the neutron-ripplon interaction in the diabatic
limit, corresponding to the ripplon frequency $\omega _{q}>\omega _{q\mathrm{%
c}}$ and the interaction potential given by Eq.\ (\ref{Vr1}), because it
deals with much larger phase space of ripplons and because, as we will see
later, the main contribution to the neutron-ripplon scattering rate comes
from the ripplons with energy $\hbar \omega _{q}\sim V_{0}\gg \hbar \omega
_{q\mathrm{c}}$.

The scattering rate $w_{\mathrm{rip}}$ of a neutron with initial in-plane
momentum $\mathbf{p}_{||}$ on ripplons is determined by two processes: the
absorption and the emission of a ripplon with wave vector $\mathbf{q}$ and
energy $\hbar \omega _{q}$,%
\begin{equation}
w_{\mathrm{rip}}=w_{\mathrm{abs}}+w_{\mathrm{em}}.
\end{equation}%
Since for typical $^{4}$He temperatures $k_{\mathrm{B}}T$ is much larger
than the initial neutron energy $K$, the populations of ripplon states with
relevant energies are $N_{q}\gtrsim 1$ or even $N_{q}\gg 1$. The phase
volume of an absorbed ripplon is much larger than that of an emitted
ripplon, because the energy of the latter is limited to the initial kinetic
energy of the neutron, $K\ll k_{\mathrm{B}}T$. Hence, one could expect that
the ripplon-neutron scattering rate is dominated by ripplon absorption, so
that $w_{\mathrm{rip}}\approx w_{\mathrm{abs}}$. However, because of a
low-energy divergence of $w_{\mathrm{rip}}$ (see below), the emission of
low-energy ripplons with energies $\hbar \omega _{q}\ll K$ may also be
important, and we therefore consider both these processes.

\subsection{Absorption of ripplons}

The absorption scattering rate $w_{\mathrm{abs}}$ of a neutron with initial
in-plane momentum $\mathbf{p}_{||}$ in the discrete vertical level with
energy $E_{0}$ is given by Fermi's golden rule, 
\begin{equation}
w_{\mathrm{abs}}=\frac{2\pi }{\hbar }\int \frac{N_{q}Sd^{2}\mathbf{p}_{q}}{%
(2\pi \hbar )^{2}}\int \frac{Sd^{2}\mathbf{p}_{\parallel }^{\prime }}{(2\pi
\hbar )^{2}}\sum_{n}\left\vert T_{\mathrm{if}}\right\vert ^{2}\delta \left(
\varepsilon -\varepsilon ^{\prime }\right) ,  \label{wa}
\end{equation}%
where $\mathbf{p}_{q}\equiv \hbar \mathbf{q}$ is the ripplon momentum, and $%
\mathbf{p}_{\parallel }^{\prime }$ and $n$ are the in-plane momentum and
out-of-plane quantum number of the final neutron state, respectively. 
\begin{equation}
N_{q}=\left[ \exp \left( \hbar \omega _{q}/k_{\mathrm{B}}T\right) -1\right]
^{-1}  \label{Nq}
\end{equation}%
is the Bose distribution function of ripplons with energy $\hbar \omega _{q}$
and with zero chemical potential. The matrix element $\left\vert T_{\mathrm{%
if}}\right\vert $ is given by Eq.\ (\ref{Tif2r}) and the initial total
energy by 
\begin{equation}
\varepsilon =\hbar \omega _{q}+\mathbf{p}_{\parallel }^{2}/2m+E_{0}.
\label{ei}
\end{equation}%
The final energy $\varepsilon ^{\prime }=\mathbf{p}_{\parallel }^{\prime
2}/2m+E_{n}$, after using the in-plane momentum conservation expressed by
the $\delta $-function in Eq.\ (\ref{Tif2r}), can be rewritten as%
\begin{equation}
\varepsilon ^{\prime }=\frac{p_{\parallel }^{2}+p_{q}^{2}+2p_{q}p_{\parallel
}\cos \phi }{2m}+E_{n},  \label{ef1}
\end{equation}%
where $\phi $ is the angle between $\mathbf{p}_{\parallel }$ and $\mathbf{p}%
_{q}$. The integration over the component $\mathbf{p}_{\parallel }^{\prime }$
of the final neutron momentum parallel to the surface cancels the $\delta $%
-function in Eq.\ (\ref{Tif2r}). After substitution of Eqs.\ (\ref{Tif2r}), (%
\ref{ei}) and (\ref{ef1}) to Eq.\ (\ref{wa}) we obtain%
\begin{eqnarray}
w_{\mathrm{abs}} &=&\int \frac{N_{q}Sp_{q}dp_{q}d\phi }{2\pi \hbar ^{3}}%
\sum_{n}\left\vert \psi _{\perp 0}\left( 0\right) \psi _{\perp n}^{\ast
}\left( 0\right) V_{0}\xi _{0q}\right\vert ^{2}  \notag  \label{wa1n} \\
&&\times \delta \left( \hbar \omega _{q}-\Delta E_{n}-\frac{%
p_{q}^{2}+2p_{q}p_{\parallel }\cos \phi }{2m}\right) ,  \label{wa1}
\end{eqnarray}%
where $\Delta E_{n}=E_{n}-E_{0}\approx E_{n}$ is the change of the
out-of-plane neutron energy after the ripplon absorption. The integration
over $\phi $\ cancels the $\delta $-function in Eq.\ (\ref{wa1}) and gives 
\begin{equation}
w_{\mathrm{abs}}=\int_{0}^{\infty }\frac{N_{q}Sp_{q}dp_{q}}{\pi \hbar ^{3}}%
\sum_{n}\frac{\left\vert \psi _{\perp 0}\left( 0\right) \psi _{\perp
n}^{\ast }\left( 0\right) V_{0}\xi _{0q}\right\vert ^{2}}{\sqrt{a^{2}-\left(
b-\Delta E_{n}\right) ^{2}}},  \label{wm1}
\end{equation}%
where $a\equiv p_{\parallel }p_{q}/m$ and $b\equiv \hbar \omega
_{q}-p_{q}^{2}/2m$.\cite{CommentI} We estimate this integral in the Appendix
C. This calculation gives the upper estimate of $w_{\mathrm{abs}}$ [see
Eqs.\ (\ref{wTup}), (\ref{wImax1}) and (\ref{wm6})] of%
\begin{equation}
w_{\mathrm{abs}}^{\mathrm{up}}\approx w_{>}^{\mathrm{up}}+w_{<}^{\mathrm{up}%
}+w_{\ll }^{\mathrm{up}}\approx 7\times 10^{-5}\times T\left[ \mathrm{K}%
\right] ~\mathrm{s}^{-1}.  \label{waF}
\end{equation}%
This corresponds to a mean neutron scattering time due to ripplon absorption
of $\tau _{\mathrm{rip}}>4$ hours even at $T=1$ K.

\subsection{Emission of ripplons}

The rate of emission of a ripplon by a surface-state neutron with momentum $%
\mathbf{p}_{\parallel }$ is given by Fermi's golden rule, similar to Eq.\ (%
\ref{wa}): 
\begin{equation}
w_{\mathrm{em}}=\frac{2\pi }{\hbar }\int \frac{N_{q}^{\prime }Sd^{2}\mathbf{p%
}_{q}}{(2\pi \hbar )^{2}}\int \frac{Sd^{2}\mathbf{p}^{\prime }}{(2\pi \hbar
)^{2}}\sum_{n}\left\vert T_{\mathrm{if}}\right\vert ^{2}\delta \left(
\varepsilon -\varepsilon ^{\prime }\right) .  \label{we}
\end{equation}%
Here, $\mathbf{p}_{q}\equiv \hbar \mathbf{q}$ is now the emitted-ripplon
momentum, $\mathbf{p}^{\prime }=\mathbf{p}_{\parallel }-\mathbf{p}_{q}$ is
the in-plane neutron momentum after emission of the ripplon, 
\begin{equation}
N_{q}^{\prime }=1+\left[ \exp \left( \hbar \omega _{q}/k_{\mathrm{B}%
}T\right) -1\right] ^{-1},  \label{N1}
\end{equation}%
is the ripplon population, $\varepsilon \approx \mathbf{p}_{\parallel
}^{2}/2m$ is the initial total energy, and%
\begin{equation*}
\varepsilon ^{\prime }=\frac{p_{\parallel }^{2}+p_{q}^{2}-2p_{q}p_{\parallel
}\cos \phi }{2m}+E_{n}+\hbar \omega _{q},
\end{equation*}%
is the final total energy. The matrix element is given by Eq.\ (\ref{Tif2r}%
). The integration over $\mathbf{p}^{\prime }$ in Eq.\ (\ref{we}) cancels
the $\delta $-function in the matrix element,%
\begin{eqnarray}
w_{\mathrm{em}} &=&\int \frac{SN_{q}^{\prime }p_{q}dp_{q}}{2\pi \hbar ^{3}}%
\sum_{n}\left\vert \psi _{\perp }\left( 0\right) \psi _{\perp }^{\prime \ast
}\left( 0\right) V_{0}\xi _{0q}\right\vert ^{2}  \notag \\
&&\times \int_{0}^{2\pi }d\phi \delta \left( \frac{p_{q}^{2}-2p_{q}p_{%
\parallel }\cos \phi }{2m}+E_{n}+\hbar \omega _{q}\right) .  \label{we1}
\end{eqnarray}%
\ The integration over the angle $\phi $ between $\mathbf{p}_{\parallel }$\
and $\mathbf{p}_{q}$\ in Eq.\ (\ref{we1}) is similar to that in the
preceding subsection in Eq.\ (\ref{wa1}) and gives%
\begin{equation}
w_{\mathrm{em}}=\int_{0}^{p_{\mathrm{em}}^{\max }}\frac{SN_{q}p_{q}dp_{q}}{%
\pi \hbar ^{3}}\sum_{n}\frac{\left\vert \psi _{\perp }\left( 0\right) \psi
_{\perp }^{\prime \ast }\left( 0\right) V_{0}\xi _{0q}\right\vert ^{2}}{%
\sqrt{a^{2}-\left( b_{1}+\Delta E_{n}\right) ^{2}}},  \label{we2}
\end{equation}%
where $a\equiv p_{\parallel }p_{q}/m$ as in the previous subsection, and $%
b_{1}=\hbar \omega _{q}+p_{q}^{2}/2m$. The integrand is real when $0\leq
\Delta E_{n}\approx E_{n}\leq a-b_{1}=p_{\parallel }p_{q}/m-\hbar \omega
_{q}-p_{q}^{2}/2m$. This can be satisfied when $a-b_{1}>0$, which for $%
p_{\parallel }^{2}/2m=100$ neV gives $p_{q}<p_{\mathrm{em}}^{\max }=\hbar q_{%
\mathrm{em}}^{\max }$ with $q_{\mathrm{em}}^{\max }\approx 6.5\times 10^{4}$
cm$^{-1}$. The maximum value of $a-b_{1}$ is $\sim 3\ \mathrm{neV}\ll
V_{0}=18.5$ neV. Hence, for the emission of ripplons, $E_{n}\ll V_{0}$, and
we may use Eqs.\ (\ref{En}), (\ref{alphan}) and (\ref{dn}). In addition,
instead of three intervals of parameters for the ripplon absorption, we only
need to consider one interval. Substituting Eqs.\ (\ref{3}), (\ref{xi0qa})
and the upper estimate of $\left\vert \psi _{\perp }^{\prime }\left(
0\right) \right\vert ^{2}\leq \left\vert \psi _{\perp 0}\left( 0\right)
\right\vert ^{2}$ to Eq.\ (\ref{we2}), we obtain an upper estimate $w_{%
\mathrm{em}}^{\mathrm{up}}$ for $w_{\mathrm{em}}$: 
\begin{eqnarray}
w_{\mathrm{em}}^{\mathrm{up}} &\approx &\int_{0}^{p_{\mathrm{em}}^{\max }}%
\frac{k_{\mathrm{B}}Tp_{q}dp_{q}}{2\pi \hbar \alpha p_{q}^{2}}\int \frac{%
dn\left\vert \psi _{\perp 0}^{2}\left( 0\right) V_{0}\right\vert ^{2}}{\sqrt{%
a^{2}-\left( b_{1}+E_{n}\right) ^{2}}}  \label{we3} \\
&=&\int \frac{k_{\mathrm{B}}Tdp_{q}}{\pi ^{2}\hbar ^{2}\alpha p_{q}}\frac{%
\left\vert \psi _{\perp 0}^{2}\left( 0\right) V_{0}\right\vert ^{2}}{g\sqrt{%
2m}}\int_{0}^{a-b_{1}}\frac{\sqrt{E_{n}}dE_{n}}{\sqrt{a^{2}-\left(
b_{1}+E_{n}\right) ^{2}}}.  \notag
\end{eqnarray}%
This integral resembles the one in Eq.\ (\ref{wm3}): the only difference is
the sign of $E_{n}$ in the denominator and, consequently, a different upper
integration limit. We may give an upper estimate of this integral by
replacing $\sqrt{E_{n}}$ by its maximum value $\sqrt{a-b_{1}}$ in the
integrand and by replacing the lower limit by $-b_{1}$ in Eq.\ (\ref{we3}).
This gives 
\begin{eqnarray*}
w_{\mathrm{em}}^{\mathrm{up}} &\approx &\int \frac{k_{\mathrm{B}}Tdp_{q}}{%
\pi ^{2}\hbar ^{2}\alpha p_{q}}\frac{\left\vert \psi _{\perp 0}^{2}\left(
0\right) V_{0}\right\vert ^{2}}{g\sqrt{2m}}\int_{-b_{1}}^{a-b_{1}}\frac{%
\sqrt{a-b_{1}}dE_{n}}{\sqrt{a^{2}-\left( b_{1}+E_{n}\right) ^{2}}} \\
&=&\int_{0}^{p_{\mathrm{em}}^{\max }}\frac{k_{\mathrm{B}}Tdp_{q}}{\pi \hbar
^{2}\alpha p_{q}}\frac{\left\vert \psi _{\perp 0}^{2}\left( 0\right)
V_{0}\right\vert ^{2}}{2g\sqrt{2m}}\sqrt{\frac{p_{\parallel }p_{q}}{m}-\hbar
\omega _{q}-\frac{p_{q}^{2}}{2m}}.
\end{eqnarray*}%
The integral converges. Neglecting $p_{q}^{2}/2m\ll \hbar \omega _{q}$ and
changing the integration variable to $\sqrt{p_{q}}$ we finally obtain%
\begin{equation}
w_{\mathrm{em}}^{\mathrm{up}}\approx \frac{k_{\mathrm{B}}T}{\pi \hbar
^{2}\alpha }\frac{\left\vert \psi _{\perp 0}^{2}\left( 0\right)
V_{0}\right\vert ^{2}}{g\sqrt{2m}}\frac{2}{3}\left( \frac{p_{\parallel }}{m}%
\right) ^{3/2}\sqrt{\frac{\rho \hbar }{\alpha \,}}.  \label{we5}
\end{equation}%
The rate of ripplon emission depends on the initial neutron momentum $%
p_{\parallel }$. At $K=p_{\parallel }^{2}/2m=100$ neV Eq.\ (\ref{we5}) gives 
\begin{equation}
w_{\mathrm{em}}^{\mathrm{up}}\approx 2\times 10^{-5}\ \mathrm{s}^{-1}\times T%
\left[ \mathrm{K}\right] .  \label{weF}
\end{equation}%
Combining Eqs.\ (\ref{waF}) and (\ref{weF}) we obtain an upper estimate for
the total scattering rate of a surface neutron in the lowest energy level $%
E_{0}$ by ripplons:%
\begin{equation}
w_{\mathrm{rip}}^{\mathrm{up}}=w_{\mathrm{abs}}^{\mathrm{up}}+w_{\mathrm{em}%
}^{\mathrm{up}}\approx 9\times 10^{-5}\ \mathrm{s}^{-1}\times T\left[ 
\mathrm{K}\right] .  \label{wrF}
\end{equation}%
This rate corresponds to a mean neutron scattering time due to the ripplons
of $\tau _{\mathrm{rip}}>3$ hours even at $T=1$ K.

\section{Other neutron scattering processes}

\subsection{Scattering of surface neutrons by bulk phonons}

The scattering of ultra-cold neutrons inside superfluid helium by bulk
phonons has been studied in Ref. \cite{Golub1979}. There, two main processes
were identified: (i) one-phonon absorption and (ii) one-phonon absorption
combined with emission of another phonon due to the cubic term in the phonon
Hamiltonian. The second process was found to dominate at low temperature,
resulting in a total scattering time of about $\tau _{\mathrm{ph0}}=100$ s
for a neutron propagating through liquid $^{4}$He at $T=1$ K. In our case of
a neutron above the He surface, both scattering processes are weakened by
the factor 
\begin{equation*}
\int_{-\infty }^{0}\psi _{\perp }^{2}\left( z\right) dz=\psi _{\perp
}^{2}\left( 0\right) /2\kappa \approx 1.16\times 10^{-7},
\end{equation*}%
because only a small part of the neutron wave function penetrates into the
liquid helium. Hence, for helium temperatures below $1$ K, the neutron
scattering time constant due to bulk phonons, $\tau _{\mathrm{ph}}\approx
\tau _{\mathrm{ph0}}2\kappa /\psi _{\perp }^{2}\left( 0\right) \gtrsim
10^{9} $ s, is extremely long and can safely be ignored.

\subsection{Scattering by surfons}

Recently, a new type of surface excitation was proposed \cite{SurStates} in
addition to the ripplons, in order to explain the temperature dependence of
the surface tension coefficient of liquid helium. These excitations, called
surfons, are He atoms in a quasistationary discrete quantum energy level
above the liquid helium surface \cite%
{SurStates,SurfonsJLTP2011,SurfonEvaporation}. The state is formed by the
combination of the van-der-Waals attractive potential of the bulk helium and
the hard-core repulsion between He atoms. Although there is so far only
indirect experimental evidence for this type of surface excitation, we
consider the neutron-surfon scattering rate to compare with the other
processes. The interaction potential is the same as for neutron interaction
with the helium vapor, but the surfons propagate only along the helium
surface. Therefore, the surfon-neutron interaction contains an additional
small factor $\sim z_{0}\psi _{\perp }^{2}\left( 0\right) \sim 3\times
10^{-5}$ due to a small overlap of the neutron and the surfon wave
functions. The activation energy of the surfon has been obtained from
fitting the temperature dependence of the surface tension coefficient of
liquid $^{4}$He to the experimental data \cite{SurfonsJLTP2011}. Its value, $%
\Delta _{\mathrm{s}0}\approx k_{\mathrm{B}}\times 2.67\ \mathrm{K}=3.7\times
10^{-16}$ erg, is significantly smaller than the evaporation energy $-\mu
=k_{\mathrm{B}}\times 7.17$ K of a $^{4}$He atom. Therefore, at low enough
temperature the neutron scattering by surfons will exceed the scattering
rate on helium vapor and must be considered for completeness.

In the calculation we can neglect the initial UCN momentum $p$ as compared
to the large surfon initial momentum $P_{\parallel }\sim \sqrt{2k_{\mathrm{B}%
}T\,M}$, similarly to our treatment of the scattering from helium vapor in
Sec. III. We also assume that the surfon in-plane kinetic energy is not
sufficient to evaporate the He atom from the surfon state after scattering.
The vertical neutron energy level $E_{n}$ may change, however, and the
out-of-plane neutron momentum may not be conserved because the helium
surface violates the spatial uniformity along the $z$-axis. The surfon
energy consists of the excitation energy $\Delta _{\mathrm{s}0}$ and of the
kinetic energy $K_{\mathrm{sur}}=\mathbf{P}_{\parallel }^{2}/2M$ of its
in-plane motion. The populations of the surfon states are approximately
given by the Boltzmann distribution, $N_{\mathrm{sur}}\left( P_{\parallel
}\right) \approx \exp \left[ -\left( \Delta _{\mathrm{s}0}+P_{\parallel
}^{2}/2M\right) /k_{\mathrm{B}}T\right] $. The calculation is described in
Appendix D and gives a very small upper estimate for the scattering rate of
neutrons on surfons: 
\begin{eqnarray}
w_{\mathrm{sur}}^{\mathrm{up}} &=&\frac{\left( Mk_{\mathrm{B}}T\right)
^{3/2}a_{\mathrm{He}}^{2}}{\hbar ^{2}m}\left\vert \psi _{\perp 0}^{2}\left(
0\right) \right\vert \frac{\sqrt{1.6\,\pi }}{5}\exp \left( \frac{-\Delta _{%
\mathrm{s}0}}{k_{\mathrm{B}}T}\right)  \notag  \label{wsur} \\
&\approx &4\times 10^{-8}\exp \left( \frac{-\Delta _{\mathrm{s}0}}{k_{%
\mathrm{B}}T}\right) \times T^{3/2}\left[ \mathrm{K}\right] ~\mathrm{s}^{-1}.
\label{wsurF}
\end{eqnarray}%
Hence, at temperatures $T>0.25$ K, the neutron scattering rate $w_{\mathrm{%
sur}}^{\mathrm{up}}$ by surfons is found to become much smaller than the
scattering rate $w_{\mathrm{vap}}$ by helium vapor given in Eq.\ (\ref{wv2}%
). However, for this and lower temperatures, the scattering by ripplons is
dominant, $w_{\mathrm{sur}}^{\mathrm{up}}\ll w_{\mathrm{rip}}^{\mathrm{up}}$%
, so that scattering of UCNs by surfons is negligibly small at any
temperature.

\begin{figure}[tbh]
\includegraphics[width=0.49\textwidth]{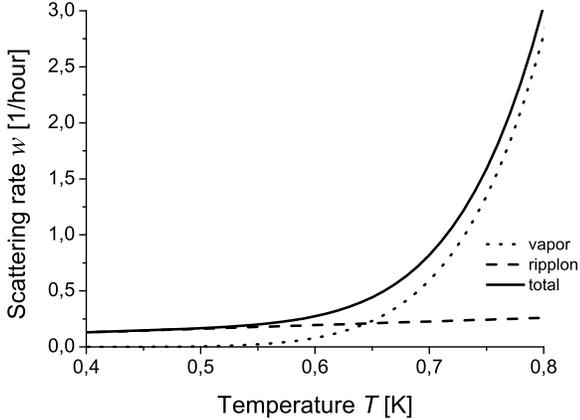}
\caption{The calculated total scattering rate (solid line) of UCNs on the
liquid helium surface in inverse hours. The dotted and dashed lines give the
contributions due to helium vapor and ripplons, respectively.}
\label{FigScatRate}
\end{figure}

Thus, the total scattering rate $w_{\mathrm{tot}}$ of UCNs on the liquid
helium surface is determined by the helium vapor at high temperatures $%
T\gtrsim 0.6$ K, and by ripplons at low temperatures $T\lesssim 0.6$ K. It
is plotted as function of temperature in Figs.\ \ref{FigScatRate} and \ref%
{FigScatRate1}.

\begin{figure}[tbh]
\includegraphics[width=0.49\textwidth]{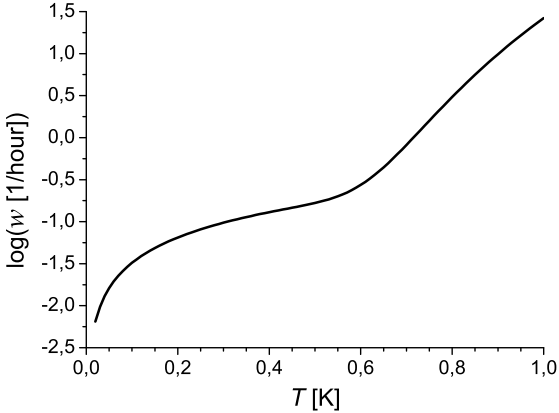}
\caption{The logarithm with base 10 of the calculated total scattering rate
of UCNs on the liquid helium surface. }
\label{FigScatRate1}
\end{figure}

\section{Discussion and sketches of experimental implementations}

The calculations presented in this paper show that, at temperatures below $%
0.7$ K, the mean scattering time of a neutron in a gravitational quantum
state above a horizontal flat surface of liquid helium is greater than the
neutron beta-decay lifetime. This surface might therefore indeed represent
an almost perfect mirror, which calls for experimental demonstration and
applications. The system could offer excellent possibilities not only to
study the quantum states represented in Fig. \ref{FigPot} but also serve as
a sensitive probe for detection of tiny energy transfers due to
helium-intrinsic or external perturbations. The application having motivated
this theoretical work is a high-precision study of the level scheme of the
neutron in the gravitational potential above the liquid mirror, giving
access to short-range, gravitation-like interactions between the neutron and
the mirror. A motivation from an experimental point of view has been current
work on new UCN sources at the ILL \cite{Zimmer/2011,Piegsa/2014} which
involves cooling many liters of ultrapure, superfluid helium below $0.7$ K.
The development had been started at the TU Munich \cite%
{Zimmer/2007,Zimmer/2010} and builds on theoretical work by Golub and
Pendlebury on superthermal UCN production via down-scattering of cold
neutrons in superfluid helium \cite{Golub/1975,Golub/1977}.

The scattering rate of neutrons on helium vapor decreases stronger than
exponentially while lowering the temperature [$\propto T^{2}\exp \left( -%
\mathrm{const}/T\right) $, see Eq.\ (\ref{wv2})]. Already for $0.8$ K it is
calculated to be smaller than the neutron decay rate. An upper estimate of
the scattering rate due to ripplons at $0.8$ K is found to be by one order
of magnitude lower than the rate due to the vapor. Owing to its linear
temperature dependence it will become the dominant contribution below about $%
0.6$ K [see Eq.\ (\ref{wrF})], however at a level already $50$ times below
the neutron decay rate. While all processes calculated here should be
insignificant for precision studies of the level scheme, experiments
involving storage of neutrons with energies up to the cutoff set by the
neutron optical potential barrier $V_{0}$ of the superfluid helium [see Eq.\
(\ref{V0})] may have different requirements. In this respect it seems
helpful that the main contribution to the neutron scattering rate at such
low temperatures is due to the low-energy part of the ripplon spectrum and
thus will dominantly lead only to transitions between nearby neutron quantum
states. Energy transfers of a few peV are however usually insufficient to
cause a neutron to penetrate through the liquid helium and thus leave the
system. Therefore, at $0.6$ K the mean escape time of a UCN with initial
kinetic energy $K<V_{0}$ could be longer than the neutron beta-decay
lifetime by several orders of magnitude. This makes an experimental set-up
using a liquid helium surface a strong candidate for a nearly loss-free
neutron container and has indeed been proposed to be applied in a neutron
lifetime experiment \cite{Bokun/1984,Alfimenkov/2009}. For highest
reliability, measurements should nonetheless be performed at different
temperatures and the container preferentially be filled with a neutron
spectrum with a gap between its upper cut-off and $V_{0}$. It should also be
noted that the value $V_{0}=18.5$\ $\mathrm{neV}$ for superfluid helium is
small compared to $\gtrsim 100$\ $\mathrm{neV}$ for conventional materials
used for neutron bottles. Counting statistics might therefore become a
limiting issue. Still, a neutron lifetime measurement employing a trap
involving a horizontal surface of superfluid helium seems an interesting
complement to projects employing magnetic neutron traps. While these possess
typical trapping potentials for low-field-seeking neutrons in the range $%
\left( 50-120\right) $\ $\mathrm{neV}$ and completely avoid any wall
collisions of truly trapped neutrons, other systematic effects such as
marginally trapped neutrons and depolarization need to be carefully
addressed \cite%
{Ezhov/2014,Salvat/2014,Leung/2014,Ezhov/2009,Leung/2009,Picker/2005,Huffman/2000,Zimmer/2000}
(see also Refs.\ \cite{Wietfeldt/2011,Paul/2009} for recent reviews and
discussion of the neutron lifetime problem).

Turning to the question how to populate and detect the neutron quantum
states above a superfluid-helium mirror (called "lake" in the sequel) one
first notes that, in contrast to a solid mirror as employed in previous and
ongoing experiments, one has to confine the liquid and deal with the
presence of a meniscus at the border of the helium container (the "coast" of
the lake). In the traditional "flow-through" scheme of current experiments
neutrons enter a mirror table with an absorbing ceiling from one side and
are detected on the other side. This might also be technically feasible for
the liquid mirror, where neutrons will then however have to enter and exit
the lake through a thin, weakly absorbing foil with low (or negative)
neutron optical potential. Application of magnetic fields might offer more
attractive, novel experimental possibilities which we sketch below.

Considering first a technique for state population, we note that one may
employ a magnetic field gradient for deceleration of neutrons moving from
above towards the horizontal mirror located at $z=0$. Neutrons with magnetic
moment $\mathbf{\mu }_{\mathrm{n}}$ in a magnetic field with modulus $B$
have a potential energy of $\pm \left\vert \mu _{\mathrm{n}}\right\vert B$,
with sign depending on the spin state with respect to the field direction.
The upper (lower) sign refers to those neutrons which become repelled
(attracted) by a positive gradient of magnetic field modulus. They are
correspondingly called low-field (high-field) seekers. Note recent
experimental demonstrations of trapped high-field seeking UCNs \cite%
{Daum/2011,Brenner/2015}. A magnetic field modulus $B\left( z\right) =Cz$,
for instance with constant (positive) gradient $C>mg/\left\vert \mu _{%
\mathrm{n}}\right\vert \approx 1.66\ \mathrm{T/m}$ overcompensates
gravitation for the high-field seeking neutrons. Those with vertical kinetic
energy $E_{\bot }=\left( \left\vert \mu _{\mathrm{n}}\right\vert C-mg\right)
h$ at height $h$ will thus have lost this energy entirely when arriving at
the mirror. This situation is analog to a neutron rising in the earth's
gravitational field to its apogee (after which it will fall back down). If
alternatively, one wants to deccelerate the low-field seeking neutrons, the
gradient has to be inverted and hence the strongest field needs to be
located at the surface. While the neutron is close to the lowest point of
its trajectory, the magnetic field gradient nearby the mirror needs to be
switched off. For limitation of the spatial region where the field needs to
be provided, one may employ a vertical, straight neutron guide, ending above
and close to the mirror. A circular absorber with a central hole for the
neutron-feeding pipe and mounted with variable distance of some tens of $%
\mathrm{\mu m}$ above the lake may serve for preparation of low neutron
quantum states as used in the first experiments \cite{Nesvizhevsky/2002}. An
obvious benefit of the magnetic population method is a possible neutron
detection acceptance angle of the full $2\pi $. Using a helium lake offers
the additional advantage that the presumably nearly perfect mirror can be
made much larger than the quartz mirrors employed in current experiments.

Compared with a flow-through experiment the sensitivity of the energy state
determination may be drastically improved using lateral UCN trapping prior
to detection, ideally for many hundreds of seconds. This might be possible
using a magnetic fence, consisting of a multipolar magnetic field
arrangement similar to the system described in Ref.\ \cite{Zimmer/2015}. For
our purpose the multipole of high order has to be oriented such as to
provide only field components within the plane defined by the helium
surface, i.e., with current-carrying rods parallel to the gravity field.
This will keep low-field seeking neutrons away from the liquid meniscus at
the container wall, thus keeping the vertical neutron state unaffected. Note
that, if one populates the lake with high-field seeking neutrons, their spin
has to be flipped after arrival at the surface and prior to arrival in the
region of strong multipolar field, which can be done, e.g., using standard
magnetic resonance techniques. Detection can still be done through the side
walls of the helium container, requiring switching-off of the magnetic
fields (or better a spin flip to turn the trapped low-field seeking neutrons
into high-field seekers to accelerate them through gaps in the magnetic
fence). Alternatively, one may let them rise back to the entrance of the
neutron guide by switching on again the magnetic field gradient used for
lake population.

Next, we discuss a possible further improvement of the lake population
technique, noting first that a vertical straight and specularly reflecting
guide with constant cross section does not mix the vertical with the
horizontal components of neutron velocity. As a result, the closer a typical
neutron approaches the mirror, the larger will be the number of reflections
per length unit of the guide. Even a small non-specularity in the reflection
will then become an issue. In addition, a typical neutron from a typical UCN
source will, after removal of its kinetic energy in the vertical direction,
still have a typical final speed parallel to the surface of several meters
per second. A lower neutron speed would however be beneficial for both the
flow-through mode and for trapping. For the former it increases the time a
neutron spends on the mirror, while for the latter a lower magnetic field
strength is sufficient for lateral UCN confinement on the lake.

A non-imaging neutron optical device proposed by Hickerson and Filippone
offers an interesting remedy for the aforementioned deficiencies of a
straight guide \cite{Hickerson/2013}. They describe a compound parabolic
concentrator (CPC) for neutrons rising from a Lambertian horizontal disk
source upwards against the gravitational field. Its neutron-optical
properties are based on the "neutron fountain" \cite{Steyerl/1988} valid for
constant force fields along the symmetry axis of a parabolic reflecting
surface. Using a constant magnetic field gradient that overcompensates the
effect of gravitation, neutrons approaching the lake from above will
experience a constant deceleration $a=C/m-g$. We may thus apply the CPC
inverted in space with the neutron source (an aperture with radius $R$)
located at height $h$ above the lake. According to the formulas given in
Ref.\ \cite{Hickerson/2013}, neutrons starting there at time $t=0$ and with
speed $v_{0}$ will, after a time $T=v_{R}/a$ (where $v_{R}=\sqrt{%
v_{0}^{2}+2aR}$) and with typically fewer than two reflections, arrive
within a narrow band of heights $0\leq z<R$ above the horizontal mirror. The
spread of total kinetic energy within the ensemble of UCN is then only $%
\Delta E\approx maR/3$ and independent of $v_{0}$. After switching off the
field gradient, the neutrons close to the mirror will thus move with much
reduced lateral velocities compared to the traditional beam method. Hence,
even without lateral trapping, state population via a CPC will lead to much
increased interaction times with the mirror and a corresponding gain in
accuracy of measurements. A CPC will be most beneficial at a pulsed UCN
source, preferably in combination with a rebunching technique as
demonstrated in Ref.\ \cite{Arimoto/2012}, and hence works best in
combination with a UCN trapping experiment. We note that the very low
lateral neutron velocities allow for quite modest magnetic trapping fields,
which makes it easy to provide large openings for neutron detection in the
magnetic fence. Obviously, a CPC with a magnetic deceleration system might
also be used for a sufficiently large conventional mirror.

In summary, this paper has given positive answers concerning necessary
prerequisites for application of a superfluid-helium mirror for study and
application of neutron quantum levels in the Earth gravity field. Further
investigations will be needed to address questions on neutron manipulation,
e.g., if transitions between levels can be induced by vibrations of the
helium surface in a controlled way. It might also be worthwhile to consider
further possibilities to create a flat mirror of large surface, such as
"Fomblin" oil (a fluorinated, organic compound with low neutron absorption,
already tested as part of an optical system for a new neutron charge
measurement \cite{Siemensen/2015}) or liquid or solid neon.

\bigskip

P. G. thanks A. M. Dyugaev for useful discussions. The work was supported by
RFBR grant \#13-02-00178.

\appendix{}

\section{Derivation of the neutron-ripplon interaction in the adiabatic
approximation}

The Schr\"{o}dinger equation for a neutron is given by%
\begin{equation}
\hat{H}\psi \left( \mathbf{r},t\right) =i\hbar \partial \psi \left( \mathbf{r%
},t\right) /\partial t,  \label{Schrod}
\end{equation}%
where the Hamiltonian 
\begin{equation}
\hat{H}=\hat{K}+\hat{V}=-\frac{\hbar ^{2}\hat{\Delta}}{2m}+mgz+V_{0}\theta %
\left[ -z+\xi (\mathbf{r}_{\parallel },t)\right]  \label{H}
\end{equation}%
contains the neutron kinetic energy $\hat{K}=-\hbar ^{2}\hat{\Delta}/2m$ and
the potential energy $\hat{V}=mgz+V_{0}\theta \left[ -z+\xi (\mathbf{r}%
_{\parallel },t)\right] $. The latter contains\ the effects of the Earth's
gravitational field and the potential wall due to the liquid helium, as
shown in Fig. \ref{FigPot}. The difference from Eq.\ (\ref{H0}) is that the
liquid helium has now a time- and space-periodic boundary $\xi (\mathbf{r}%
_{\parallel },t)$ given by Eq.\ (\ref{xi2}). The difference between Eq.\ (%
\ref{H}) and Eq.\ (4) from Ref. \cite{Abele/2010} is that the surface has
now a periodic spatial dependence.

The adiabatic adjustment of the neutron wave function to the new surface
profile means that the neutron wave function, in first approximation,
adiabatically shifts in $z$-direction by the length $\xi (\mathbf{r}%
_{\parallel },t)$: $\psi \left( \mathbf{r},t\right) \rightarrow \tilde{\psi}%
\left( \mathbf{r}+\xi (\mathbf{r}_{\parallel },t)\boldsymbol{\hat{z}}%
,t\right) $, where $\boldsymbol{\hat{z}}$ is the unitary vector in $z$%
-direction. This shift can be written via the translation ($z$-shift)
operator 
\begin{equation*}
\hat{T}_{z}\left( \xi \right) =\exp \left[ \xi (\mathbf{r}_{\parallel
},t)\partial /\partial z\right] =\exp \left[ i\xi (\mathbf{r}_{\parallel
},t)p_{z}/\hbar \right] .
\end{equation*}%
Its action on the wave function is%
\begin{equation*}
\hat{T}_{z}\left( \xi \right) \psi _{\perp }\left( z\right) =\psi _{\perp
}\left( z+\xi \right) .
\end{equation*}%
We also define a new wave function%
\begin{equation*}
\tilde{\psi}\left( \mathbf{r}+\xi \boldsymbol{\hat{z}},t\right) =\psi \left( 
\mathbf{r},t\right) =\hat{T}_{z}\left( \xi \right) \tilde{\psi}\left( 
\mathbf{r},t\right) ,
\end{equation*}%
which after substitution into Eq.\ (\ref{Schrod}) gives a new Schr\"{o}%
dinger equation for $\tilde{\psi}\left( \mathbf{r},t\right) $:%
\begin{equation}
\hat{H}\hat{T}_{z}\left( \xi \right) \tilde{\psi}\left( \mathbf{r},t\right)
=i\hbar \partial \left( \hat{T}_{z}\left( \xi \right) \tilde{\psi}\left( 
\mathbf{r},t\right) \right) /\partial t.  \label{NSchrod}
\end{equation}%
The action of the shift operator on the potential energy function $V\left( 
\mathbf{r}\right) $ is given by 
\begin{equation}
V\left( \mathbf{r}\right) \hat{T}_{z}\left( \xi \right) =\hat{T}_{z}\left(
\xi \right) V\left( \mathbf{r}-\xi (\mathbf{r}_{\parallel },t)\boldsymbol{%
\hat{z}}\right) ,  \label{Vs}
\end{equation}%
while for the commutator with kinetic-energy operator $\hat{K}=-\hbar ^{2}%
\hat{\Delta}/2m$ we have%
\begin{gather}
\hat{K}\hat{T}_{z}\left( \xi \right) -\hat{T}_{z}\left( \xi \right) \hat{K}=-%
\frac{\hbar ^{2}}{m}\left( \boldsymbol{\nabla }^{2}e^{\xi (\mathbf{r}%
_{\parallel },t)\partial /\partial z}-e^{\xi (\mathbf{r}_{\parallel
},t)\partial /\partial z}\boldsymbol{\nabla }^{2}\right)  \notag \\
=\frac{2\hat{p}_{\parallel }\hat{p}_{q}+\hat{p}_{q}^{2}}{2m}\xi (\mathbf{r}%
_{\parallel },t)\frac{\partial }{\partial z}=\frac{\left( \hat{p}_{\parallel
}+\hat{p}_{q}\right) ^{2}-\hat{p}_{\parallel }^{2}}{2m}\xi (\mathbf{r}%
_{\parallel },t)\frac{\partial }{\partial z},  \label{Cr0}
\end{gather}%
where $\hat{p}_{\parallel }=-i\hbar \boldsymbol{\nabla }_{\parallel }$ and $%
\hat{p}_{q}=\hbar q$ are the neutron and the ripplon momentum operators
along the surface, respectively. The time-dependence of $\xi (\mathbf{r}%
_{\parallel },t)$ also gives an additional term on the r.h.s. of Eq.\ (\ref%
{NSchrod}):%
\begin{gather}
i\hbar \frac{\partial }{\partial t}\left( \hat{T}_{z}\tilde{\psi}\left( 
\mathbf{r},t\right) \right) =i\hbar \frac{\hat{T}_{z}\partial \tilde{\psi}%
\left( \mathbf{r},t\right) }{\partial t}+i\hbar \frac{\partial \hat{T}_{z}}{%
\partial t}\tilde{\psi}\left( \mathbf{r},t\right)  \notag \\
=i\hbar \frac{\hat{T}_{z}\partial \tilde{\psi}\left( \mathbf{r},t\right) }{%
\partial t}+\hbar \omega _{q}\xi (\mathbf{r}_{\parallel },t)\frac{\partial }{%
\partial z}\hat{T}_{z}\tilde{\psi}\left( \mathbf{r},t\right) .  \label{Cr1}
\end{gather}%
Combining Eqs.\ (\ref{NSchrod}) and (\ref{Cr1}) we obtain a new Schr\"{o}%
dinger equation, 
\begin{equation}
\hat{T}_{z}\left\{ \hat{H}_{0}+\hat{H}_{\mathrm{int}}-i\hbar \partial
/\partial t\right\} \tilde{\psi}\left( \mathbf{r},t\right) =0,
\end{equation}%
where $\hat{H}_{0}$ is given by Eq.\ (\ref{H0}) and the interaction term is
given by%
\begin{equation}
\hat{H}_{\mathrm{int}}=\xi (\mathbf{r}_{\parallel },t)\left\{ \left[ \frac{2%
\hat{p}_{\parallel }\hat{p}_{q}+\hat{p}_{q}^{2}}{2m}-\hbar \omega _{q}\right]
\frac{\partial }{\partial z}+mg\right\} .  \label{Hint0}
\end{equation}

\section{Crossover between adiabatic and diabatic limits in classical physics%
}

For a classical particle above the surface in the limit $V_{0}\rightarrow
\infty $ the crossover between diabatic and adiabatic limits occurs when the
maximal acceleration of the helium surface $\partial ^{2}\xi _{q}/\partial
t^{2}=\omega _{q}^{2}\xi _{q}$, due to its oscillatory motion, becomes equal
to the free fall acceleration $g$, 
\begin{equation}
\omega _{q\mathrm{c}}^{2}\xi _{q}=g.  \label{ClCr}
\end{equation}%
The classical amplitude $\xi _{q}$ of the surface oscillations with wave
vector $\mathbf{q}$ differs from $\xi _{0q}$\ in Eq.\ (\ref{xi0qa}) by the
square root of the Bose distribution function $N_{q}$ given by Eq.\ (\ref{Nq}%
):\cite{CommentNq} 
\begin{equation}
\xi _{q}=\sqrt{N_{q}}\xi _{0q}\approx \xi _{0q}\sqrt{k_{\mathrm{B}}T/\hbar
\omega _{q}}.  \label{xiq}
\end{equation}%
In addition, $\xi _{0q}$\ in Eq.\ (\ref{xi0qa}) depends on the surface $S$,
which must be defined. In the formulas for the neutron scattering rate by
ripplons this surface-dependence is unphysical and does not occur
explicitely, because the $S$-dependence of the ripplon amplitude in Eq.\ (%
\ref{xi0qa}) is compensated by the $S$-dependence of the ripplon density of
states (see below). Similarly, the total mean square amplitude of thermal
surface oscillations at any point $r_{\parallel }$ is given by the sum over
all $\mathbf{q}$-vectors,%
\begin{equation*}
\left\langle \xi ^{2}\left( r_{\parallel }\right) \right\rangle =\sum_{%
\mathbf{q}}\xi _{\mathbf{q}}^{2}=\int N_{q}\xi _{0\mathbf{q}}^{2}\frac{Sd^{2}%
\mathbf{q}}{\left( 2\pi \right) ^{2}},
\end{equation*}%
and the surface area $S$ drops out. More generally, if we are interested in
the surface waves with the wave number $q$ in some interval $\Delta
q_{x}\Delta q_{y}$, then we sum all ripplon modes in the phase volume $%
S\Delta q_{x}\Delta q_{y}$, and the surface area $S$ again drops out from
the total $\left\langle \xi ^{2}\right\rangle $. In the estimate (\ref{ClCr}%
) for the diabatic-adiabatic crossover the surface $S$ is defined by the
area $S_{\mathrm{n}}$ of the neutron wave function along the surface, which
corresponds to the momentum smearing $\Delta q_{x}\Delta q_{y}\sim S_{%
\mathrm{n}}^{-1}$.

For a lower estimate of $\omega _{q\mathrm{c}}$, leading to an upper
estimate of the ripplon scattering rate, we take the minimal possible $S$
given by the square of the wave length: $S_{\min }\approx \left( 2\pi
/q\right) ^{2}$. Then, substituting it to Eq.\ (\ref{xi0qa}), we have 
\begin{equation}
\xi _{0q}^{\max }\approx \left( \frac{\hbar q^{2}}{8\pi ^{2}\,\sqrt{\rho
\alpha q}}\right) ^{1/2}=\frac{q^{3/4}}{2\pi }\left( \frac{\hbar /2}{\sqrt{%
\rho \alpha }}\right) ^{1/2}.  \label{xiL}
\end{equation}%
For $q=1\ \mathrm{\mu m}^{-1}=10^{4}$ cm$^{-1}$ this formula gives $\xi
_{0q}^{\max }\approx 7\times 10^{-5}$ nm, which is much less than $q^{-1}$.
The corresponding 
\begin{equation}
\xi _{q}^{\max }\approx \xi _{0q}^{\max }\sqrt{\frac{k_{\mathrm{B}}T}{\hbar
\omega _{q}}}=\frac{1}{2\pi }\left( \frac{k_{\mathrm{B}}T}{2\alpha }\right)
^{1/2}\approx 0.02\ \mathrm{nm}\sqrt{T\,\text{[K]}}
\end{equation}%
is also much less than $q^{-1}$, and we can apply the usual surface wave
description. Substituting $S_{\min }\approx \left( 2\pi /q\right) ^{2}$ to
Eqs.\ (\ref{xi0qa}), (\ref{ClCr}) and (\ref{Nq}) gives%
\begin{equation*}
g=\omega _{q\mathrm{c}}^{3/2}\left( \frac{k_{\mathrm{B}}T\,q^{3/2}}{2\left(
2\pi \right) ^{2}\,\sqrt{\rho \alpha }}\right) ^{1/2},
\end{equation*}%
which, using Eq.\ (\ref{3}), gives the lowest estimate for the crossover
frequency%
\begin{equation}
\omega _{q\mathrm{c}}=\sqrt{2\pi g\sqrt{\frac{2\,\alpha \,}{k_{\mathrm{B}}T}}%
}\approx \frac{6.6\times 10^{5}\ \text{s}^{-1}}{\left( T\,\text{[K]}\right)
^{1/4}}.  \label{omcCl}
\end{equation}%
This frequency corresponds to the neutron energy (at $T=1$ K) $\hbar \omega
_{q\mathrm{c}}\approx 7\times 10^{-22}\ \mathrm{erg}\approx 0.44\ \mathrm{neV%
}>E_{0}$ and to 
\begin{equation}
q_{\mathrm{c}}=0.56\ \mathrm{\mu m}^{-1}>\varkappa .  \label{qcCl}
\end{equation}%
Hence, in the diabatic limit $q>q_{\mathrm{c}}$, and we can always use the
ripplon dispersion given by Eq.\ (\ref{3}).

Another condition of the classical adiabatic limit is that the curvature of
the surface $\nabla ^{2}\xi =q^{2}\xi _{q}^{\max }$ is less than the
curvature of the neutron trajectory due to the parabolic free-fall motion $%
\partial ^{2}z/\partial r_{\parallel }^{2}=g/v_{\parallel }^{2}$, where $%
v_{\parallel }$ is the neutron velocity along the surface. Taking a UCN
kinetic energy of $K_{\parallel }=100$ neV, corresponding to $v_{\parallel
}^{2}=2K_{\parallel }/m=19$ m$^{2}$/s$^{2}$, we can check that the condition 
$q^{2}\xi _{q}^{\max }<g/v_{\parallel }^{2}$ is fulfilled at $q=q_{\mathrm{c}%
}=0.56\ \mathrm{\mu m}^{-1}$. Hence, the condition $\omega _{q}<\omega _{q%
\mathrm{c}}$ given by Eq.\ (\ref{omcCl}) ensures the classical adiabatic
limit.

\section{Calculations for the neutron scattering rate due to ripplons}

In this section we evaluate the integral in Eq.\ (\ref{wa1}) or (\ref{wm1}),
which gives the neutron scattering rate by ripplons. The integration over $%
p_{q}$ and $n$ in Eqs.\ (\ref{wa1}) can be separated into several regions,
given by different limits of the ratio $p_{q}/p_{\parallel }$ and of the
difference $\Delta E_{n}-V_{0}$. For $\Delta E_{n}<V_{0}$ the final neutron
vertical state belongs to a discrete energy spectrum, approximately given by
Eq.\ (\ref{En}). For $\Delta E_{n}>V_{0}$ the final neutron vertical state
belongs to the continuous energy spectrum and can approximately be taken as
a plane wave.

In the region $p_{q}\gg p_{\parallel }$ the initial neutron kinetic energy
is negligible and, for the majority of the scattering events, the change of
the neutron out-of-plane kinetic energy $\Delta E_{n}>V_{0}$. The integral
in Eq.\ (\ref{wa1}) is evaluated in this limit in Appendix C1 below.

In the region of small ripplon momentum, $p_{q}\lesssim p_{\parallel }$,
studied in Appendix C2, the angle $\phi $ between the initial neutron and
ripplon momenta is important for the out-of-plane energy transfer $\Delta
E_{n}$, and the scattering rate depends on the initial neutron momentum $%
p_{\parallel }$. Depending on the sign of the difference $\Delta E_{n}-V_{0}$%
, this region is split into two. For $\Delta E_{n}<V_{0}$ the final neutron
state belongs to the discrete spectrum and is described by the formulas in
Sec. II. For $\Delta E_{n}>V_{0}$ the final vertical neutron state belongs
to the continuous spectrum and can be approximated by Eqs.\ (\ref{psinpT})
and (\ref{EnT}).

\subsection{Absorption of thermal (high-energy) ripplons}

In this subsection we consider the region of large momenta $p_{q}\gg
p_{\parallel }=\sqrt{2Km}$ contributing to the integral in Eq.\ (\ref{wa1}).
Let us assume the in-plane kinetic energy $K$ of ultra-cold neutrons being
less than $K_{\ast }=100$ neV, which corresponds to a maximal initial
neutron wave number $q_{\ast }=p_{\parallel \ast }/\hbar =7\times 10^{5}\ 
\mathrm{cm}^{-1}=0.07$ nm$^{-1}$ and to a maximal neutron velocity $%
v_{\parallel \ast }=\sqrt{2K/m}=4.4$ m/s. For $q=q_{\ast }$ the ripplon
energy, according to Eq.\ (\ref{3}), is given by 
\begin{equation}
\hbar \omega _{q\ast }\equiv \hbar \omega _{q}\left( \hbar q=p_{\parallel
\ast }\right) \approx 600\ \mathrm{neV}\approx 7\ \mathrm{mK}\gg
V_{0},K_{\ast },  \label{oms}
\end{equation}%
and the ripplon velocity is $v_{q\ast }=3\omega _{q\ast }/2q_{\ast }=20$ m/s 
\cite{CommentQT}. If a ripplon with such a high energy is absorbed, the
final out-of-plane neutron energy $E_{n}\sim \hbar \omega _{q}$ is much
higher than the potential barrier $V_{0}=18.5$ neV. It is then reasonable to
take the final out-of-plane neutron wave function as a plane wave, 
\begin{equation}
\psi _{\perp n}\left( z\right) \approx \exp \left( ip_{z}^{\prime }z/\hbar
\right) /\sqrt{L_{z}}.  \label{psinpT}
\end{equation}%
Accordingly, the neutron out-of-plane energy can be approximated by the
free-particle quadratic dispersion 
\begin{equation}
E_{n}\approx p_{z}^{\prime 2}/2m,  \label{EnT}
\end{equation}%
where $p_{z}^{\prime }$ is the component of the final neutron momentum
perpendicular to the surface. The sum over out-of-plane neutron wave number $%
n$ in Eq.\ (\ref{wa1}) then becomes an integral over $p_{z}^{\prime }$: 
\begin{equation}
\sum_{n}\rightarrow \int \rho _{n}\left( p_{z}^{\prime }\right)
dp_{z}^{\prime },  \label{SumN}
\end{equation}%
where the one-dimensional neutron density of states is given by \cite%
{CommentSumN} 
\begin{equation}
\rho _{n}\left( p_{z}^{\prime }\right) \approx L_{z}/2\pi \hbar .
\label{DoSn}
\end{equation}

For scattering by thermal ripplons with $q>q_{\ast }$ the initial neutron
energy $K\approx p_{\parallel }^{2}/2m$ and the momentum $p_{\parallel
}<p_{q}$ can be neglected. Eqs.\ (\ref{ei}) and (\ref{ef1}) then simplify to 
\begin{equation}
\varepsilon \approx \hbar \omega _{q};\qquad \varepsilon ^{\prime }\approx
\left( p_{q}^{2}+p_{z}^{\prime 2}\right) /2m.  \label{efa}
\end{equation}%
Using Eq.\ (\ref{xi0q}), we rewrite Eq.\ (\ref{wa1}) as (the lower index "$>$%
" means large ripplon energy):%
\begin{eqnarray}
w_{>} &=&\int_{\hbar q_{\ast }}^{\infty }\frac{\hbar \left\vert \psi _{\perp
0}\left( 0\right) V_{0}\right\vert ^{2}}{2\sqrt{\rho \alpha q}}\frac{%
N_{q}p_{q}dp_{q}}{2\pi \hbar ^{2}}  \notag \\
&&\times \int_{0}^{\infty }\frac{dp_{z}^{\prime }}{\hbar ^{2}}\delta \left(
\hbar \omega _{q}-\frac{p_{q}^{2}+p_{z}^{\prime 2}}{2m}\right)  \notag \\
&=&\int_{q_{\ast }}^{\infty }\frac{\left\vert \psi _{\perp 0}\left( 0\right)
V_{0}\right\vert ^{2}}{2\pi \hbar \sqrt{\rho \alpha }}\frac{mN_{q}\sqrt{q}dq%
}{\sqrt{2m\hbar \omega _{q}-\hbar ^{2}q^{2}}}.  \label{wr}
\end{eqnarray}%
The square root in the denominator is real at $2m\hbar \omega _{q}=2m\hbar 
\sqrt{\alpha /\rho }q^{3/2}>\hbar ^{2}q^{2}$, which gives$~q<4q_{0}\equiv
\left( 2m/\hbar \right) ^{2}\alpha /\rho \approx 2.5$ nm$^{-1}$ and
corresponds to the ripplon energy $\hbar \omega _{q}<\hbar \omega _{q\max
}\approx \hbar \left( 2m/\hbar \right) ^{3}\left( \alpha /\rho \right)
^{2}=2\times 10^{-16}\ \mathrm{erg}=1.25\times 10^{-4}\ \mathrm{eV}\approx
k_{\mathrm{B}}\times 1.5\ \mathrm{K}\gtrsim k_{\mathrm{B}}T$. Above this
energy the simple absorption of a ripplon by a UCN in a surface state is
impossible because of the conservation laws for energy and momentum.
Substituting Eqs.\ (\ref{3}) and (\ref{Nq}) to Eq.\ (\ref{wr}), and
introducing the dimensionless variable $\zeta \equiv \hbar \omega _{q}/k_{%
\mathrm{B}}T=\hbar \sqrt{\alpha /\rho }q^{3/2}/k_{\mathrm{B}}T$, for which $%
q=\left( \zeta k_{\mathrm{B}}T\sqrt{\rho /\alpha }/\hbar \right) ^{2/3}$,\
we obtain%
\begin{equation}
w_{>}=\frac{\left\vert \psi _{\perp 0}\left( 0\right) V_{0}\right\vert ^{2}%
\sqrt{mk_{B}T}}{3\pi \hbar ^{2}\alpha \sqrt{2}}\int_{\zeta _{\min }}^{\zeta
_{\max }}\frac{\zeta ^{-1/2}d\zeta ~\left( e^{\zeta }-1\right) ^{-1}}{\sqrt{%
1-\left( \zeta /\zeta _{\max }\right) ^{1/3}}},  \label{wr1}
\end{equation}%
where $\zeta _{\min }=\hbar \omega _{q\ast }/k_{\mathrm{B}}T$ is given by
Eq.\ (\ref{oms}) and $\zeta _{\max }=\hbar \omega _{q\max }/k_{\mathrm{B}%
}T=\left( 2m\right) ^{3}\left( \alpha /\hbar \rho \right) ^{2}/k_{\mathrm{B}%
}T\sim 1$. The integration in Eq.\ (\ref{wr1}) diverges as $\zeta _{\min
}^{-1/2}$ at the lower limit, and the main part of the integral comes from
this divergence:%
\begin{equation}
w_{>}\approx \frac{\left\vert \psi _{\perp 0}\left( 0\right)
V_{0}\right\vert ^{2}\sqrt{mk_{\mathrm{B}}T}}{3\pi \hbar ^{2}\alpha \sqrt{2}%
\sqrt{\zeta _{\min }}}.  \label{wra}
\end{equation}%
Substituting the cutoff $\zeta _{\min }=\hbar \omega _{q\ast }/k_{\mathrm{B}%
}T$ given by Eq.\ (\ref{oms}) and other numerical values to Eq.\ (\ref{wra}%
), we obtain the contribution to the neutron scattering rate from the
high-energy ripplons with $\hbar q>p_{\parallel \ast }$: 
\begin{equation}
w_{>}\approx 1.7\times 10^{-6}\ \mathrm{s}^{-1}\times T\left[ \mathrm{K}%
\right] .  \label{wm}
\end{equation}%
At smaller ripplon energy, i.e. at $\hbar q<p_{\parallel \ast }$, the
integral in Eq.\ (\ref{wa1}) must be estimated without the approximation in
Eqs.\ (\ref{psinpT})-(\ref{efa}). In the next subsection we show that Eq.\ (%
\ref{wra}) overestimates the integral in Eq.\ (\ref{wa1}) for $\hbar
q<p_{\parallel }$, especially for $\hbar \omega _{q}\lesssim V_{0}$ where
the infrared divergence disappears.

At $K_{\ast }\rightarrow 0$, when the cutoff given by $q_{\ast }=\hbar
/p_{\parallel \ast }$ is too small, the infrared divergence in Eq.\ (\ref%
{wra}) must be cut off at $\zeta _{\min }\approx V_{0}/k_{\mathrm{B}}T$,
because the approximation given by Eqs.\ (\ref{psinpT})-(\ref{DoSn}) is not
valid for lower ripplon energies, for which the neutron state after the
absorption still belongs to the discrete spectrum along the $z$-axis. A
rough estimate of the absorption rate of high-energy ripplons with $\hbar
\omega _{q}>V_{0}$ can be obtained for small initial neutron energies $%
K<V_{0}$ by substituting $\zeta _{\min }\approx V_{0}/k_{\mathrm{B}}T$ to
Eq.\ (\ref{wra}): 
\begin{equation}
w_{>}^{\mathrm{up}}\approx w_{>}\left( \zeta _{\min }\approx V_{0}/k_{%
\mathrm{B}}T\right) \approx 10^{-5}\ \mathrm{s}^{-1}\times T\left[ \mathrm{K}%
\right] .  \label{wTup}
\end{equation}%
This estimate gives a neutron mean scattering time $1/w_{>}^{\mathrm{up}%
}\approx 27$ hours, which is much greater than the intrinsic neutron
lifetime.

\subsection{Upper estimate of the absorption rate of low-energy ripplons}

For $k_{\mathrm{B}}T\gg \hbar \omega _{q}$ the ripplon population is given
by $N_{q}\approx k_{\mathrm{B}}T/\hbar \omega _{q}$. Substituting Eqs.\ (\ref%
{3}) and (\ref{xi0qa}) to Eq.\ (\ref{wm1}) we obtain%
\begin{equation}
w_{<}=\int_{0}^{p_{\max }}\frac{k_{\mathrm{B}}Tdp_{q}^{2}}{4\,\pi \hbar
\alpha p_{q}^{2}}\sum_{n}\frac{\left\vert \psi _{\perp 0}\left( 0\right)
\psi _{\perp n}^{\ast }\left( 0\right) V_{0}\right\vert ^{2}}{\sqrt{%
a^{2}-\left( b-E_{n}\right) ^{2}}}.  \label{wm2}
\end{equation}

\subsubsection{Transitions to a continuous neutron spectrum}

In this subsection we consider the case of final neutron energies $%
E_{n}>V_{0}$ above the potential barrier and thus belonging to a continuous
spectrum. We may then apply the approximation given by Eqs.\ (\ref{psinpT})-(%
\ref{DoSn}) and rewrite Eq.\ (\ref{wm2}) as 
\begin{equation}
w_{<}\approx \int_{p_{\min }}^{p_{\max }}\frac{k_{\mathrm{B}}Tdp_{q}^{2}}{%
4\pi \hbar \alpha p_{q}^{2}}\frac{\left\vert \psi _{\perp 0}\left( 0\right)
V_{0}\right\vert ^{2}}{2\pi \hbar }I,  \label{wmi}
\end{equation}%
where the integral 
\begin{equation}
I\equiv \int \frac{dp_{z}^{\prime }}{\sqrt{a^{2}-\left( b-p_{z}^{\prime
2}/2m\right) ^{2}}}=\int_{V_{0}}^{a+b}\frac{\sqrt{m/2E_{n}}dE_{n}}{\sqrt{%
a^{2}-\left( b-E_{n}\right) ^{2}}}.  \label{I1ab}
\end{equation}%
For $b>a$ we may give an upper estimate of this integral: 
\begin{equation}
I<I_{\max }=\frac{\sqrt{m}}{\sqrt{2V_{0}}}\int_{b-a}^{b+a}\frac{dE_{n}}{%
\sqrt{a^{2}-\left( b-E_{n}\right) ^{2}}}=\frac{\pi \sqrt{m}}{\sqrt{2V_{0}}}.
\label{I2ab}
\end{equation}%
The corresponding upper estimate of Eq.\ (\ref{wmi}) is 
\begin{equation}
w_{<}^{\mathrm{up}}\approx \frac{k_{\mathrm{B}}T\left\vert \psi _{\perp
0}\left( 0\right) V_{0}\right\vert ^{2}\sqrt{m}}{4\pi \hbar ^{2}\alpha \sqrt{%
2V_{0}}}\ln \left( \frac{p_{\max }}{p_{\min }}\right) .  \label{wImax}
\end{equation}%
The interval of integration $V_{0}\leq E_{n}\leq b+a$ in Eq.\ (\ref{I1ab})
is nonzero for $b+a\approx \hbar \omega _{q}+p_{\parallel }p_{q}/m>V_{0}$,
which for $p_{\parallel }^{2}/2m=100$ neV corresponds to $q>q_{\min }\equiv
p_{\min }/\hbar \approx 4\times 10^{4}$ cm$^{-1}$. Substituting also $\psi
_{\perp 0}\left( 0\right) \approx 0.236$ cm$^{-1/2}$ and $p_{\max
}=p_{\parallel \ast }$ to Eq.\ (\ref{wImax}), we obtain%
\begin{equation}
w_{<}^{\mathrm{up}}\approx 7.3\times 10^{-6}\ln \left( q_{\ast }/q_{\min
}\right) \times T\left[ \mathrm{K}\right] \approx 2\times 10^{-5}\times T%
\left[ \mathrm{K}\right] .  \label{wImax1}
\end{equation}

\subsubsection{Transitions to the discrete neutron levels}

In this subsection we consider the case of final neutron energies in the
interval $0<\Delta E_{n}\lesssim V_{0}$ below the potential barrier and
approximately given by Eqs.\ (\ref{En}) and (\ref{alphan}). Since $V_{0}\gg
E_{0}$, the sum over $n$ in Eq.\ (\ref{wm2}) still includes many terms and
can be approximated by an integration over $n$ for $n\gg 1$. Eqs.\ (\ref{En}%
) and (\ref{alphan}) give $E_{n}\approx mgz_{0}\left( 3\pi n/2\right) ^{2/3}$%
, which can be rewritten as%
\begin{equation*}
n\approx \frac{2}{3\pi }\left( \frac{\Delta E_{n}}{mgz_{0}}\right) ^{3/2}=%
\frac{2\left( \Delta E_{n}\right) ^{3/2}}{3\pi g\hbar }\sqrt{\frac{2}{m}}
\end{equation*}%
and gives%
\begin{equation}
\frac{dn}{dE_{n}}=\frac{\sqrt{E_{n}}}{\pi g\hbar }\sqrt{\frac{2}{m}}.
\label{dn}
\end{equation}%
We also use that $\left\vert \psi _{\perp n}^{\ast }\left( 0\right)
\right\vert \lesssim \psi _{\perp 0}\left( 0\right) $, and for an upper
estimate of Eq.\ (\ref{wm2}) we replace $\left\vert \psi _{\perp n}^{\ast
}\left( 0\right) \right\vert $ by $\left\vert \psi _{\perp 0}\left( 0\right)
\right\vert $ for $E_{n}<V_{0}$. Using Eq.\ (\ref{dn}), we rewrite Eq.\ (\ref%
{wm2}) for $\Delta E_{n}<V_{0}$ as%
\begin{equation}
w_{\ll }\approx \int_{0}^{p_{\max }}\frac{k_{\mathrm{B}}Tdp_{q}}{\pi \hbar
^{2}\alpha p_{q}}\frac{\left\vert \psi _{\perp 0}^{2}\left( 0\right)
V_{0}\right\vert ^{2}}{\pi g\sqrt{2m}}\int_{0}^{V_{0}}\frac{\sqrt{E_{n}}%
dE_{n}}{\sqrt{a^{2}-\left( b-E_{n}\right) ^{2}}}.  \label{wm3}
\end{equation}%
This integral converges, with main contributions from $E_{n}\sim V_{0}$. The
upper estimate $w_{\ll }^{\mathrm{up}}$ of this integral can be obtained by
replacing $\sqrt{E_{n}}$ by $\sqrt{V_{0}}$ in the integrand and by extending
the integration region from $\left( 0,V_{0}\right) $ to $\left(
a-b,a+b\right) $. This gives an integral over $E_{n}$ similar to Eq.\ (\ref%
{I2ab}): 
\begin{eqnarray}
w_{\ll }^{\mathrm{up}} &=&\int_{p_{\min }}^{p_{\max }}\frac{k_{\mathrm{B}%
}Tdp_{q}}{\pi \hbar ^{2}\alpha p_{q}}\frac{\left\vert \psi _{\perp
0}^{2}\left( 0\right) V_{0}\right\vert ^{2}\sqrt{V_{0}}}{\pi g\sqrt{2m}} 
\notag \\
&&\times \int_{b-a}^{b+a}\frac{dE_{n}}{\sqrt{a^{2}-\left( b-E_{n}\right) ^{2}%
}}  \notag \\
&=&\frac{k_{B}T\left\vert \psi _{\perp 0}^{2}\left( 0\right)
V_{0}\right\vert ^{2}\sqrt{V_{0}}}{\pi \hbar ^{2}\alpha g\sqrt{2m}}\ln
\left( \frac{p_{\max V0}}{p_{\min V0}}\right) .  \label{wm5}
\end{eqnarray}%
Since $V_{0}>E_{n}$ and the integrand in Eq.\ (\ref{wm3}) is real for $%
b-a<E_{n}<b+a$, the region of integration over $E_{n}$ in Eq.\ (\ref{wm3})
is nonzero if $b-a\approx \hbar \omega _{q}-p_{\parallel }p_{q}/m<V_{0}$,
which for $p_{\parallel }^{2}/2m=100$ neV corresponds to $p_{q}<p_{\max
}\approx \hbar q_{\max V0}$ with $q_{\max V0}=1.57\times 10^{5}$ cm$^{-1}$.
On the other hand, $\Delta E_{n}<b+a$ can reach $V_{0}$ if $b+a\approx \hbar
\omega _{q}+p_{\parallel }p_{q}/m\geq V_{0}$. For $p_{\parallel }^{2}/2m=100$
neV this gives $p_{q}>p_{\min V0}\approx \hbar q_{\min V0}$ with $q_{\min
V0}=3.8\times 10^{4}$ cm$^{-1}$. For $p_{q}<p_{\min V0}$ the logarithmic
divergence disappears. Hence, using Eq.\ (\ref{wm5}) we obtain an upper
estimate of $w_{\ll }$:%
\begin{equation}
w_{\ll }^{\mathrm{up}}\approx 4\times 10^{-5}\times T\left[ \mathrm{K}\right]
~\mathrm{s}^{-1}.  \label{wm6}
\end{equation}

\section{Scattering of neutrons by surfons}

\subsection{Matrix element}

A surfon, being a $^{4}$He atom on the surface energy level \cite%
{SurStates,SurfonsJLTP2011}, interacts with a neutron via the potential
given in Eq.\ (\ref{Vi}). The matrix elements of neutron-He interaction in
Eqs.\ (\ref{T1}),(\ref{Tif2}) assumes that the He wave function is a plane
wave also along the $z$-axis, which is not the case for the surfons.
Therefore, in this subsection we derive the neutron-surfon matrix element in
the way similar to that in Sec. IIIA.

The surfon wave function is given by a product $\Psi =\Psi _{\perp }\left(
z\right) \Psi _{\parallel }\left( \mathbf{r}_{\parallel }\right) $. The
parallel-to-surface surfon wave function is a plane wave: $\Psi _{\parallel
}\left( \mathbf{r}_{\parallel }\right) =\exp \left( i\mathbf{P}_{\parallel
}\,\mathbf{r}_{\parallel }/\hbar \right) $. The perpendicular-to-surface
surfon wave function $\Psi _{\perp }\left( z\right) $ was analyzed in Refs. 
\cite{SurfonEvaporation,SurfonsJLTP2011,SurfonMobility} and shown to be
localized above the surface on a height $\sim 0.4\ \mathrm{nm}\ll z_{0}$ 
\cite{CommentSurfon}. Hence, for our calculation we may take $\Psi _{\perp
}^{2}\left( z\right) \approx \delta \left( z\right) $. The initial and final
surfon wave functions differ only by the initial and final momentum, $%
\mathbf{P}_{\parallel }$ and $\mathbf{P}_{\parallel }^{\prime }$,
respectively. On the other hand, the neutron out-of-plane wave function
strongly changes due to the scattering on a surfon, and for the majority of
events it gets transferred from the discrete to the continuous spectrum.
Hence, the final neutron wave function is given by Eq.\ (\ref{psiNf}). The
matrix element of the interaction potential (\ref{Vi}) is given by\emph{\ } 
\begin{eqnarray}
T_{\mathrm{if}} &=&U\int d^{3}\mathbf{r}\boldsymbol{\,}\psi _{\perp 0}\left(
z\right) \psi _{\parallel }\left( \mathbf{r}_{||}\right) \psi ^{\prime
}\left( \mathbf{r}\right) \times  \notag \\
&&\int d^{3}\mathbf{R}\psi \Psi _{\perp }^{2}\left( z\right) \exp \left[ 
\frac{i\left( \mathbf{P}_{\parallel }-\mathbf{P}_{\parallel }^{\prime
}\right) \boldsymbol{\,}\mathbf{R}_{\parallel }}{\hbar }\right] \,\delta
^{(3)}\left( \mathbf{r}-\mathbf{R}\right)  \notag \\
&\approx &\,U\psi _{\perp 0}\left( 0\right) \left( 2\pi \hbar \right)
^{2}\delta ^{(2)}\left( \Delta \mathbf{P}_{\mathrm{tot}||}\right) /\sqrt{SV},
\label{Ts}
\end{eqnarray}%
where $\Delta \mathbf{P}_{\mathrm{tot}}\approx \mathbf{P}-\mathbf{P}^{\prime
}-\mathbf{p}^{\prime }$ is the change of total momentum and $\mathbf{R}$ is
the surfon coordinate. Below we need only the square of the absolute value
of the matrix element $T_{\mathrm{if}}$. The square of the $\delta $%
-function in $\left\vert T_{\mathrm{if}}\right\vert ^{2}$\ should be treated
using Eq.\ (\ref{delta2}). Then instead of Eq.\ (\ref{Tif2}) we obtain 
\begin{equation}
\left\vert T_{\mathrm{if}}\right\vert ^{2}=\,U^{2}\psi _{\perp 0}^{2}\left(
0\right) \left( 2\pi \hbar \right) ^{2}\delta ^{(2)}\left( \Delta \mathbf{P}%
_{\mathrm{tot}||}\right) /V.  \label{Ts2}
\end{equation}

\subsection{Scattering rate}

The scattering rate of a neutron by a surfon with initial momentum $\mathbf{P%
}_{\parallel }$ is given by the square of the matrix element (\ref{Tif2})
integrated over the final momenta $\mathbf{p}^{\prime }$ and $\mathbf{P}%
_{\parallel }^{\prime }$ of the neutron and surfon, respectively (Fermi's
golden rule \cite{LL3}), 
\begin{equation}
w_{\mathbf{p}}=\frac{2\pi }{\hbar }\int \frac{d^{2}\mathbf{P}_{\parallel
}^{\prime }}{(2\pi \hbar )^{2}}\int \frac{Vd^{3}\mathbf{p}^{\prime }}{(2\pi
\hbar )^{3}}\left\vert T_{\mathrm{if}}\right\vert ^{2}\delta \left(
\varepsilon -\varepsilon ^{\prime }\right) .  \label{ws1}
\end{equation}%
Here $\varepsilon \approx P_{\parallel }^{2}/2M$ and $\varepsilon ^{\prime
}=P_{\parallel }^{\prime 2}/2M+p^{\prime 2}/2m$ are the initial and final
total energies of surfon + neutron, respectively. We now substitute them and
Eq.\ (\ref{Ts2}) to Eq.\ (\ref{ws1}). The integration over $\mathbf{p}%
_{\parallel }^{\prime }$ cancels $\delta ^{(2)}\left( \Delta \mathbf{P}_{%
\mathrm{tot}||}\right) $ in the matrix element in Eq.\ (\ref{Ts2}), where\ $%
\Delta \mathbf{P}_{\mathrm{tot}||}\approx \mathbf{P}_{\parallel }-\mathbf{P}%
_{\parallel }^{\prime }-\mathbf{p}^{\prime }$: 
\begin{equation}
w_{\mathbf{p}}=\frac{U^{2}\left\vert \psi _{\perp 0}^{2}\left( 0\right)
\right\vert }{\hbar ^{3}}\int \frac{P_{\parallel }^{\prime }dP_{\parallel
}^{\prime }}{2\pi }\frac{dp_{\perp }^{\prime }}{2\pi \hbar }\int d\phi
\,\delta \left( \varepsilon -\varepsilon ^{\prime }\right) .  \label{ws2}
\end{equation}%
Substituting $\varepsilon $ and $\varepsilon ^{\prime }$, and integrating
over the angle $\phi $ between $\mathbf{P}_{\parallel }$ and $\mathbf{P}%
_{\parallel }^{\prime }$, we obtain%
\begin{gather}
w_{\mathbf{p}}=\frac{U^{2}}{\hbar ^{3}}\left\vert \psi _{\perp 0}^{2}\left(
0\right) \right\vert \int \frac{P_{\parallel }^{\prime }dP_{\parallel
}^{\prime }}{2\pi }\frac{dp_{\perp }^{\prime }}{2\pi \hbar }\times
\label{ws2m} \\
\left[ \left( \frac{P_{\parallel }^{\prime }P_{\parallel }}{m}\right)
^{2}-\left( \frac{P_{\parallel }^{2}}{2M}-\frac{P_{\parallel }^{\prime 2}}{2M%
}-\frac{P_{\parallel }^{2}+P_{\parallel }^{\prime 2}+p_{\perp }^{\prime 2}}{%
2m}\right) ^{2}\right] ^{-1/2}.  \notag
\end{gather}%
Now we use $M=4m$ to simplify this expression: 
\begin{equation*}
w_{\mathbf{p}}=\int \frac{\left( U^{2}m\left\vert \psi _{\perp 0}^{2}\left(
0\right) \right\vert /4\pi ^{2}\hbar ^{4}\right) ~~P_{\parallel }^{\prime
}dP_{\parallel }^{\prime }dp_{\perp }^{\prime }}{\sqrt{P_{\parallel
}^{\prime 2}P_{\parallel }^{2}-\left[ P_{\parallel }^{2}/8-P_{\parallel
}^{\prime 2}/8-\left( P_{\parallel }^{2}+P_{\parallel }^{\prime 2}+p_{\perp
}^{\prime 2}\right) /2\right] ^{2}}}.
\end{equation*}%
Introducing new dimensionless integration variables $x_{p}\equiv
P_{\parallel }^{\prime 2}/P_{\parallel }^{2}$ and $y_{p}\equiv p_{\perp
}^{^{\prime }}/P_{\parallel }$, we rewrite this as 
\begin{equation}
w_{\mathbf{p}}=\frac{U^{2}mP_{\parallel }}{\pi ^{2}\hbar ^{4}}\int \frac{%
\left\vert \psi _{\perp 0}^{2}\left( 0\right) \right\vert ~dx_{p}~dy_{p}}{%
\sqrt{64x_{p}-\left( 3+5x_{p}+4y_{p}^{2}\right) ^{2}}}.  \label{ws3}
\end{equation}%
This may be further transformed to%
\begin{eqnarray}
w_{\mathbf{p}} &=&\frac{U^{2}mP_{\parallel }}{\pi ^{2}\hbar ^{4}}\int \frac{%
\left\vert \psi _{\perp 0}^{2}\left( 0\right) \right\vert dx_{p}~dy_{p}}{%
\sqrt{\left( y_{p}^{2}-b_{1}\right) \left( b_{2}-y_{p}^{2}\right) }}
\label{ws3m} \\
&=&\frac{U^{2}mP_{\parallel }}{\pi ^{2}\hbar ^{4}}\int \frac{\left\vert \psi
_{\perp 0}^{2}\left( 0\right) \right\vert dx_{p}~dy_{p}}{5\sqrt{\left(
x_{p}-a_{1}\right) \left( a_{2}-x_{p}\right) }},  \label{ws3x}
\end{eqnarray}%
where the solutions of square-root equations are%
\begin{equation}
b_{1,2}=\frac{-3\mp 8\sqrt{x_{p}}-5x_{p}}{4},  \label{b12}
\end{equation}%
and%
\begin{equation}
a_{1,2}=\frac{17\mp 8\sqrt{1-20y_{p}^{2}}-20y_{p}^{2}}{25}.  \label{a12}
\end{equation}%
The integral over $x_{p}$ in Eq.\ (\ref{ws3x}) gives $\pi $ for any $%
a_{2}>a_{1}$, however, the integrand is real only for some values of $%
y_{p}^{2}$. The maximum value of $b_{2}$ in Eq.\ (\ref{b12}) is $b_{2}^{\max
}=0.2$ at $x_{p}=16/25$. Since $0<y_{p}^{2}<b_{2}^{\max }$, for an upper
esimate of the integral in Eq.\ (\ref{ws3}) we may take 
\begin{eqnarray}
w_{\mathbf{p}}^{\mathrm{up}} &=&\frac{U^{2}mP_{\parallel }}{\pi ^{2}\hbar
^{4}}\int \frac{\left\vert \psi _{\perp 0}^{2}\left( 0\right) \right\vert
~dx_{p}~2\sqrt{b_{2}^{\max }}}{5\sqrt{\left( x_{p}-a_{1}\right) \left(
a_{2}-x_{p}\right) }}  \notag \\
&=&\frac{U^{2}mP_{\parallel }}{\pi ^{2}\hbar ^{4}}\left\vert \psi _{\perp
0}^{2}\left( 0\right) \right\vert \frac{2\pi \sqrt{0.2}}{5}.  \label{wsup}
\end{eqnarray}%
The total scattering rate is given by the integral over all initial surfon
states with corresponding populations:%
\begin{equation*}
\frac{w_{\mathrm{sur}}^{\mathrm{up}}}{\left\vert \psi _{\perp 0}^{2}\left(
0\right) \right\vert }=\int \frac{d^{2}\mathbf{P}_{\parallel }}{(2\pi \hbar
)^{2}}w_{\mathbf{p}}^{\mathrm{up}}\exp \left( \frac{-\Delta _{\mathrm{s}0}-%
\mathbf{P}_{\parallel }^{2}/2M}{k_{\mathrm{B}}T}\right) .
\end{equation*}%
Substituting Eq.\ (\ref{wsup}) and performing the integration we obtain%
\begin{equation*}
\frac{w_{\mathrm{sur}}^{\mathrm{up}}}{\left\vert \psi _{\perp 0}^{2}\left(
0\right) \right\vert }=\frac{U^{2}m\sqrt{1.6\pi }}{20\pi ^{2}\hbar ^{6}}\exp
\left( \frac{-\Delta _{\mathrm{s}0}}{k_{\mathrm{B}}T}\right) \left( Mk_{%
\mathrm{B}}T\right) ^{3/2}.
\end{equation*}%
Substituting $U$ from Eq.\ (\ref{Vi}) we obtain the scattering rate given by
Eq.\ (\ref{wsur}), which is negligibly small.


\begin{thebibliography}{99}
\bibitem{CommentBoltzmann} Since the chemical potential of liquid helium $%
\left\vert \mu \right\vert \gg k_{\mathrm{B}}T$, the Boltzmann distribution
does not differ from the Bose-Einstein distribution.
\bibitem{CommentSW} Eq.\ (\ref{xi0q}) can be obtained\cite{Shikin} by
equating the energy of a classical surface wave with the wave number $q$ and
amplitude $\xi _{0q}$ on the area $S$, given in \S 12,25,62 of Ref. \cite%
{LL6}, to the energy $\hbar \omega _{q}$ of one ripplon.
\bibitem{CommentNq} The thermally excited ripplons are not coherent,
therefore the population factor $N_{q}$ increases the mean energy of the
surface wave by $N_{q}$\ times, not the amplitude of thermal ripplons, which
is increased only by $\sqrt{N_{q}}$ times.
\bibitem{CommentQT} The typical thermal ripplon has even larger energy $%
\hbar \omega _{q}\sim k_{\mathrm{B}}T\approx 0.5$ K, corresponding to the
wave number $q\approx 1.2$ nm$^{-1}$ and the velocity $v_{q}\equiv \partial
\omega _{q}/\partial q\approx 82$ m/s.
\bibitem{CommentSumN} Eqs.\ (\ref{psinpT})-(\ref{SumN}) can also be applied
for smaller $E_{n}$ if $p_{z}^{\prime }=p_{z}^{\prime }\left( z\right) $ is
understood as a quasi-classical coordinate-dependent momentum; then in Eq.\ (%
\ref{EnT}) one then should take its value at $z=0$, and $\rho _{n}\left(
p_{z}^{\prime }\right) $ differs from Eq.\ (\ref{DoSn}) and depends on the
actual spectrum.
\bibitem{CommentI} The integrand in Eq.\ (\ref{wm1}) is real at $b-a\leq
\Delta E_{n}\leq b+a.$The quantity $b>a$ at $p_{\parallel }<m\sqrt{\alpha
q/\rho }-\hbar q/2$, which for $p_{\parallel }^{2}/2m=100$ neV corresponds
to $q>p_{\parallel }^{2}\rho /m^{2}\alpha \approx 8\times 10^{4}$ cm$^{-1}$
and $\hbar \omega _{q}>23$ neV.
\end{thebibliography}
\end{document}